\documentclass[12pt]{article}

\usepackage[mathscr]{eucal}
\usepackage{epsfig,amsfonts}
\usepackage{amsmath}
\usepackage{amsthm,amssymb}
\usepackage{bm}
\usepackage{mathbbol} 
\usepackage{graphicx}
\usepackage{relsize}
\usepackage{hhline}
\usepackage{comment}
\usepackage{bbm}
\usepackage[dvipsnames]{xcolor}
\usepackage{printlen}
\usepackage{cite}
\usepackage{psfrag}
\usepackage{mathrsfs} 
\usepackage{hyperref}

\usepackage{array}
\usepackage{tikz}
\usepackage{enumerate}
\usepackage{textcomp}

\usepackage{braket}
\usepackage{float}
\usepackage{enumitem}
\usepackage{subcaption}
\usepackage{ulem}
\usepackage{arydshln}
\numberwithin{equation}{section}

\newcommand{\BR}[2]{u^{[#1]}_{#2}}

\newcommand{\ri}{{\rm i}}

\definecolor{gold(metallic)}{rgb}{0.83, 0.69, 0.22}
\definecolor{gold(web)(golden)}{rgb}{1.0, 0.84, 0.0}

\newcommand{\be}{\begin{eqnarray}}
	\newcommand{\ee}{\end{eqnarray}}
\newcommand{\non}{\nonumber}
\newcommand{\id}{\mathbb{I}}

\newcommand{\tr}{\mathop{\rm tr}\nolimits}
\newcommand{\diag}{\mathop{\rm diag}\nolimits}

\begin{document}
	
	\begin{titlepage}
		$\phantom{I}$
		\vspace{1.8cm}
		
		\begin{center}
			\begin{LARGE}
				{Quantum-group-invariant $D^{(2)}_{n+1}$ models: Bethe ansatz and finite-size spectrum}
			\end{LARGE}
			\vspace{1cm}
			\begin{large}
				{\bf 
					
					Holger Frahm$^1$, Sascha Gehrmann$^{1,2}$,\\ Rafael I. Nepomechie$^3$, Ana L. Retore$^{4,5}$}
				
			\end{large}
			
			\vspace{0.8cm}
			$^1$Institut f$\ddot{{\rm u}}$r Theoretische Physik, 
			Leibniz Universit$\ddot{{\rm a}}$t Hannover\\
			Appelstra\ss e 2, 30167 Hannover, Germany\\\vspace{0.4cm}
			$^2$The Rudolf Peierls Centre for Theoretical Physics,\\ Oxford
			University, Oxford OX1 3PU, UK\\ \vspace{0.4cm}
			$^3$Department of Physics, PO Box 248046\\
			University of Miami, Coral Gables, FL 33124 USA\\ \vspace{0.4cm}
			$^4$School of Mathematics $\&$ Hamilton Mathematics Institute,\\ Trinity College Dublin, Ireland\\ \vspace{0.5cm}
			$^5$Department of Mathematical Sciences, Durham University\\ Durham, DH1 3LE, UK

			\vspace{0.8cm}
			
		\end{center}
		
		\begin{center}
			
			\parbox{16cm}{%
				\centerline{\bf Abstract} \vspace{.5cm}
				We consider the quantum integrable spin chain models associated with the Jimbo R-matrix based on the quantum affine algebra $D^{(2)}_{n+1}$, subject to quantum-group-invariant boundary conditions parameterized by two discrete variables $p=0,\dots, n$ and $\varepsilon = 0, 1$.  
				We develop the analytical Bethe ansatz for the previously unexplored case $\varepsilon = 1$ with any $n$, and use it to investigate the effects of different boundary conditions on the finite-size spectrum of the quantum spin chain based on the rank-$2$ algebra $D^{(2)}_3$.
				Previous work on this model with periodic boundary conditions has shown that it is critical for the range of anisotropy parameters $0<\gamma<\pi/4$, where its scaling limit is described by a non-compact CFT with continuous degrees of freedom related to two copies of the 2D black hole sigma model.  The scaling limit of the model with quantum-group-invariant boundary conditions depends on the parameter $\varepsilon$:
				similarly as in the rank-$1$ $D^{(2)}_2$ chain, we find that the symmetry of the lattice model is spontaneously broken, and the spectrum of conformal weights has both discrete and continuous components, for $\varepsilon=1$. For $p=1$, the latter coincides with that of the $D^{(2)}_2$ chain, which should correspond to a non-compact brane related to one black hole CFT in the presence of boundaries. For $\varepsilon=0$, the spectrum of conformal weights is purely discrete.
				
			}
			
		\end{center}
		
		\vfill
		
	\end{titlepage}

	\section{Introduction}
	
	The fascinating discovery in 1989 by Pasquier and Saleur \cite{Pasquier:1989kd} of an open quantum spin chain that is both integrable and quantum-group-invariant led to a long-running search for more such models and their Bethe ans\"atze, see e.g. \cite{Mezincescu:1990hda,Nepomechie:2018dsn, Nepomechie:2018wzp, Nepomechie:2018nvl} and references therein. 
	The bulk interactions of these models are dictated by solutions (so-called R-matrices) \cite{Bazhanov:1984gu, Jimbo:1985ua, Bazhanov:1986mu}
	of the Yang-Baxter equation that are associated with affine Lie algebras $\hat g$, while the boundary interactions are prescribed by certain corresponding solutions (so-called K-matrices) \cite{Mezincescu:1990hda, Batchelor:1996np, Malara:2004bi, Martins:2000xie, Nepomechie:2018wzp} of the boundary Yang-Baxter equation \cite{Cherednik:1984vvp, Sklyanin:1988yz, Ghoshal:1993tm}.
	The Bethe ansatz for corresponding models with periodic boundary conditions (BCs), and therefore without quantum-group symmetry, was found by Reshetikhin \cite{Reshetikhin:1987}. Roughly speaking,
	the Bethe equations for the quantum-group-invariant models \cite{Nepomechie:2018nvl} (depending on a discrete parameter $p$) are ``doubled'' versions of those found by Reshetikhin, with an additional factor corresponding to deleting the $p$-th node from the $\hat g$ 
	Dynkin diagram.\footnote{There exist more general K-matrices, which lead to models without quantum-group symmetry and with more complicated Bethe ans\"atze, see e.g. \cite{Li:2019rzy, Li:2021qeq, Li:2022clv, Lu:2025tef} and references therein.}
	
	Among the R-matrices associated with infinite families of
	affine Lie algebras, those associated with $D^{(2)}_{n+1}$ \cite{Jimbo:1985ua} are the most complicated. It is therefore not surprising that, apart from \cite{Martins:2000xie}, there had been until recently relatively little work on models constructed with these R-matrices. Interest in these models grew when it was realized \cite{Frahm:2012eb, Robertson:2020eri} that the $D^{(2)}_{2}$ R-matrix can (roughly speaking) be factorized into a product of four $A^{(1)}_{1}$ R-matrices, implying that $D^{(2)}_{2}$ models are related to lattice models (staggered 6-vertex model, antiferromagnetic (afm) Potts model) that have non-compact degrees of freedom \cite{Ikhlef:2008zz, Ikhlef:2011ay, Frahm:2013cma}.
	
	More precisely, the integrable boundary conditions for the quantum-group-invariant $D^{(2)}_{n+1}$ models in \cite{Nepomechie:2018wzp} depend on a parameter $\varepsilon$ that can take values 0 or 1. For the simplest case $n=1$ (that is, $D^{(2)}_{2}$), it was found \cite{Robertson:2020imc, FrGe22} that continuous degrees of freedom appear only for $\varepsilon=1$, which opens up possibilities for studying aspects of the $D$-brane constructions for non-compact boundary CFTs \cite{RiSc04,Scho06,CreHikRon2011} starting from a lattice model. It is then natural to ask what happens for $n>1$; and the present work represents a first step towards addressing this question. (As a warm-up for this challenging problem, we investigated the finite-size spectrum of the quasi-periodic $D^{(2)}_{3}$ model in \cite{Frahm:2023lpe}.)
	
	Since the Bethe ansatz for the $D^{(2)}_{n+1}$ models with $\varepsilon=1$ was heretofore not known except for $n=1$ \cite{Nepomechie:2019tbr, Robertson:2020imc, Nepomechie:2020onv}, we begin in Section \ref{sec:general} by reviewing the construction and symmetries of these models, and by presenting their analytical Bethe ansatz solution. 
	In Section~\ref{sec:D22}, we briefly recall known results from the literature for the simplest case \( n = 1 \). We then focus in Section~\ref{sec:D23}
	on the case \( n = 2 \). We introduce the Hamiltonians for four 
	distinct quantum-group-invariant boundary conditions, and present the main numerical results of our finite-size analysis. Finally, in Section~\ref{sec:discussion}, we conclude with a brief discussion of our findings, along with several conjectures and open problems.

	\section{The $D^{(2)}_{n+1}$ models}\label{sec:general}
	
	We consider integrable open quantum spin chains constructed \cite{Cherednik:1984vvp, Sklyanin:1988yz}  with the $D^{(2)}_{n+1}$ R-matrix \cite{Jimbo:1985ua}, $n=1, 2, \ldots$, and with corresponding K-matrices \cite{Nepomechie:2018wzp} that depend on two discrete parameters: $p$ (which can take $n+1$ different values, namely, $p=0, 1, \ldots, n$) and $\varepsilon$ (which can take two different values, namely, $\varepsilon = 0, 1$). The transfer matrices for these spin chains have quantum group (QG) symmetry $U_{q}(B_{n-p}) \otimes U_{q}(B_{p})$, as well as a $p \leftrightarrow n-p$ duality symmetry.
	
	The eigenvalues of the transfer matrix and the corresponding
	Bethe equations for the models with $\varepsilon = 0$ 
	were proposed in \cite{Nepomechie:2018nvl}.  We add here  
	the corresponding results for the cases with $\varepsilon = 1$, for all possible values of $n$ and $p$, which had previously not been reported.
	
	For the case $n=1$ with $\varepsilon = 1$, a Bethe ansatz that accounts for part of the spectrum was proposed in \cite{Nepomechie:2019tbr}.
	A modification of this Bethe ansatz that could account for the the complete spectrum was proposed in
	\cite{Robertson:2020imc}, and was subsequently proved 
	(using the factorization \cite{Robertson:2020eri} of the R-matrix) by algebraic Bethe ansatz in \cite{Nepomechie:2020onv}. Our Bethe ansatz for general $n$ with $\varepsilon = 1$ 
	reduces for $n=1$ to the one in 
	\cite{Robertson:2020imc, Nepomechie:2020onv}.  The
	continuum limit of this model is described \cite{Robertson:2020imc, FrGe22, FrGK24} by 
	a {\it non-compact} CFT 
	and is briefly reviewed in Section \ref{sec:D22}. 
	
	After briefly reviewing the construction of the transfer matrix and its symmetries in Sec. \ref{subsec:transf}, we present our proposed Bethe ansatz for general values of $n, \varepsilon, p$ in Sec. \ref{subsec:BA}.
	
	\subsection{The transfer matrix and its symmetries}\label{subsec:transf}
	
	We consider the $D^{(2)}_{n+1}$ R-matrix $R(u)$
	(solution of the Yang-Baxter equation) given by \eqref{RD2},
    with spectral parameter $u$ and anisotropy parameter $\eta$. 
	The right K-matrix $K^{R}(u, \varepsilon, p)$ (solution of the boundary Yang-Baxter equation \cite{Cherednik:1984vvp, Ghoshal:1993tm})
	is the block-diagonal matrix
	given by \cite{Nepomechie:2018wzp}
	\begin{align}
		{\renewcommand{\arraystretch}{1.2}
			K^{R}(u, \varepsilon, p) = \left(\begin{array}{c;{2pt/2pt}c;{2pt/2pt}c c;{2pt/2pt}c;{2pt/2pt}c}
				k_{-}(u)\mathbb{I}_{p\times p} & & &  & &\\ \hdashline[2pt/2pt]
				& g(u)\mathbb{I}_{(n-p)\times (n-p)} & & & &  \\ \hdashline[2pt/2pt]
				& & k_1(u) & k_2(u) & &\\
				& & k_2(u) & k_1(u) & &\\ \hdashline[2pt/2pt]
				& & & & g(u)\mathbb{I}_{(n-p)\times (n-p)} &\\ \hdashline[2pt/2pt]
				& & & & & k_{+}(u)\mathbb{I}_{p\times p}
			\end{array}\right),
			\nonumber} \\ 
		\label{KRb}
	\end{align}
	\noindent
	where $p= 0, 1, \ldots, n $, and 
	\begin{align}
		k_{\pm}(u) &=e^{\pm 2u}\,, \nonumber\\
		g(u) &=\frac{\cosh\left(u-(n-2p)\eta+\frac{\ri 
				\pi}{2}\varepsilon\right)}{\cosh\left(u+(n-2p)\eta-\frac{\ri 
				\pi}{2}\varepsilon\right)}\,, \nonumber\\
		k_1(u) &=\frac{\cosh(u)\cosh\left((n-2p)\eta+\frac{\ri 
				\pi}{2}\varepsilon\right)}{\cosh\left(u+(n-2p)\eta+\frac{\ri 
				\pi}{2}\varepsilon\right)}\,, \nonumber\\
		k_2(u) &=-\frac{\sinh(u)\sinh\left((n-2p)\eta+\frac{\ri 
				\pi}{2}\varepsilon\right)}{\cosh\left(u+(n-2p)\eta+\frac{\ri 
				\pi}{2}\varepsilon\right)} \,,
	\end{align}
	with $\varepsilon = 0, 1$. 
	The left K-matrix 
	$K^{L}(u, \varepsilon, p)$ is given by 
	\begin{equation}
		K^{L}(u, \varepsilon, p) = K^{R}(-u-\rho, \varepsilon, p)\, M\,, \qquad \rho = - 2 n \eta\,, 
		\label{KL}
	\end{equation}
	where the matrix $M$ is given by \eqref{Mmat},
	which corresponds to imposing the ``same'' boundary conditions on the two ends.
	
	The transfer matrix for an integrable open quantum spin chain of length $N$, with bulk and boundary interactions dictated by these R- and K- matrices respectively, 
	is given by \cite{Sklyanin:1988yz} 
	\be
	t(u, \varepsilon, p) = \tr_a K^{L}_{a}(u, \varepsilon, p)\, T_a(u)\,  K^{R}_{a}(u, \varepsilon ,p)\, \widehat{T}_a(u) \,, 
	\label{transfer}
	\ee
	where the single-row monodromy matrices are defined by
	\begin{align} 
		T_a(u) &= R_{aN}(u)\ R_{a N-1}(u)\ \cdots R_{a1}(u) \,,  \non \\
		\widehat{T}_a(u) &= R_{1a}(u)\ \cdots R_{N-1 a}(u)\ R_{Na}(u) \,,  
		\label{monodromy}
	\end{align}
	and the trace in (\ref{transfer}) is over the ``auxiliary'' space, which is denoted by $a$. The dimension of the local Hilbert space at each site is $2n+2$.
	The transfer matrix is engineered to satisfy the commutativity property
	\be
	\left[ t(u, \varepsilon ,p) \,, t(v, \varepsilon ,p) \right] = 0 \,,
	\label{commutativity} 
	\ee
	and contains the Hamiltonian and higher local conserved quantities. The transfer matrix is 
	also crossing invariant
	\be
	t(u, \varepsilon ,p) = t(-u-\rho, \varepsilon ,p)\,.
	\label{transfercrossing}
	\ee
	
\begin{figure}[H] 
		\centering
		\begin{tikzpicture}
			\node at (0,0){\includegraphics[width=0.9\textwidth]{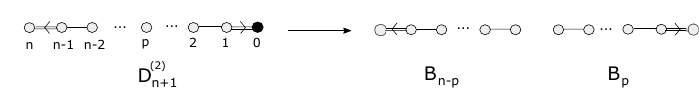}};
			\draw[red] (-4.44,0.65)--(-4.44,0.15);
		\end{tikzpicture}
		\caption{Subalgebras of $D^{(2)}_{n+1}$ corresponding to removing the $p$-th node from the extended Dynkin diagram. The ``affine node'' is black and labeled 0.
			\label{fig:Dynkin}  }
	\end{figure}

	As discussed in \cite{Nepomechie:2018wzp, Nepomechie:2018nvl}, the transfer matrix $t(u, \varepsilon ,p)$
	has the QG symmetry $U_{q}(B_{n-p}) \otimes 
	U_{q}(B_{p})$, corresponding to cutting and removing the $p$-th node from the $D^{(2)}_{n+1}$ Dynkin diagram, as shown in Figure \ref{fig:Dynkin}.
	
	\noindent
	Hence, for $0 < p < n$, the QG symmetry is given by a tensor product of two factors, to which we refer as the ``left'' ($U_{q}(B_{n-p})$) and ``right'' ($U_{q}(B_{p})$) factors.
	For $p=0$, the ``right'' factor is absent; while for 
	$p=n$, the ``left'' factor is absent.
	That is,
	\begin{align}
		& \left[\Delta_{N}(H_{i}^{(\ell)}(p)) \,, 
		t(u, \varepsilon,p) \right] = 
		\left[\Delta_{N}(E_{i}^{\pm (\ell)}(p)) \,, 
		t(u, \varepsilon, p) \right] = 0 
		\,, \qquad i = 1\,, \ldots \,, n-p\,, \non \\
		& \left[\Delta_{N}(H_{i}^{(r)}(p)) \,, 
		t(u, \varepsilon, p) \right] = 
		\left[\Delta_{N}(E_{i}^{\pm (r)}(p)) \,, 
		t(u, \varepsilon ,p) \right] = 0 
		\,, \qquad i = 1\,, \ldots \,, p \,,
		\label{QGsym}
	\end{align}
	where $H_{i}^{(\ell)}(p)$ and $E_{i}^{\pm (\ell)}(p)$ are Cartan and raising/lowering operators of the 
	``left'' algebra $B_{n-p}$\,; $H_{i}^{(r)}(p)$ and 
	$E_{i}^{\pm (r)}(p)$ are corresponding generators of the ``right'' algebra $B_{p}$\,;
	and $\Delta_{N}$ denotes the $N$-fold 
	coproduct, see Appendix \ref{subap:generators} for details.
	
	The transfer matrix also has a $p \leftrightarrow n-p$ ``duality'' symmetry that exchanges the ``left'' and ``right'' factors
	\be
	{\cal U}\, t(u, \varepsilon, p)\, {\cal U}^{-1} = f(u,\varepsilon ,p)\, t(u, \varepsilon, n-p)  \,, 
	\label{duality}
	\ee 
	with
	\be
	f(u,\varepsilon, p) = \frac{\cosh(u-(n+2p)\eta + \frac{\ri\pi}{2}\varepsilon)}
	{\cosh(u-(3n-2p)\eta - \frac{\ri\pi}{2}\varepsilon)}
	\frac{\cosh(u-(n-2p)\eta + \frac{\ri\pi}{2}\varepsilon)}
	{\cosh(u+(n-2p)\eta - \frac{\ri\pi}{2}\varepsilon)} \,,
	\ee
	see Appendix \ref{subap:dualityandgenerators} for the definition of ${\cal U}$. 
	In particular, for $p=\frac{n}{2}$ ($n$ even),
	the transfer matrix is self-dual
	\begin{equation}
		\left[ {\cal U}\,,  t(u, \varepsilon, p=\tfrac{n}{2}) \right] = 0 \,,
		\label{selfdual}
	\end{equation}
	since $f(u, \varepsilon, \frac{n}{2}) = 1$.

	For $p=\frac{n}{2}$ ($n$ even) and
	$\varepsilon=1$, there is an additional (``bonus'') symmetry, which leads to even higher degeneracies for the transfer-matrix eigenvalues \cite{Nepomechie:2018wzp}.
	
	Finally, the transfer matrix has the $Z_2$ symmetry 
	\begin{equation}
		t(u,\varepsilon, p) = \mathbb{U}^{\otimes N}\,  t(u,\varepsilon, p)\,  \mathbb{U}^{\otimes N} \,, 
		\qquad \mathbb{U} = \left(\begin{array}{c;{2pt/2pt}c c;{2pt/2pt}c}
			\id_{n \times n} & & &  \\ \hdashline[2pt/2pt]
			& 0 & 1 &   \\
			& 1 & 0 &   \\ \hdashline[2pt/2pt]
			& & &  \id_{n\times n}
		\end{array}\right) \,,
		\label{Z2symmetry}
	\end{equation}
	as for the closed chain \cite{Frahm:2023lpe}.
	
	\subsection{Analytical Bethe ansatz}\label{subsec:BA}
	
	We present here the generalization of the analytical Bethe ansatz from \cite{Nepomechie:2018nvl}, which was for $\varepsilon = 0$, to both values $\varepsilon = 0, 1$.  It turns out that the analysis is very similar to the one in \cite{Nepomechie:2018nvl}, except for a key difference  in \eqref{Qln}  below.  We therefore closely follow the latter reference, and simply present the results.
	
	\subsubsection{Eigenvalues of the transfer matrix}
	
	The transfer matrix and Cartan generators can be diagonalized 
	simultaneously 
	\begin{align}
		t(u, \varepsilon, p)\, |\Lambda^{(m_{1}, \ldots\,, m_{n})}(\varepsilon, p)\rangle &= \Lambda^{(m_{1}, 
			\ldots, m_{n})}(u, \varepsilon, p)\,
		|\Lambda^{(m_{1}, \ldots, m_{n})}(\varepsilon, p)\rangle \,, \non \\
		\Delta_{N}(H_{i}^{(\ell)}(p))\, |\Lambda^{(m_{1}, \ldots, m_{n})}(\varepsilon, p)\rangle &= 
		h^{(\ell)}_{i}\,
		|\Lambda^{(m_{1}, \ldots, m_{n})}(\varepsilon, p)\rangle \,, \qquad i = 1, \ldots, 
		n-p \,, \non \\
		\Delta_{N}(H_{i}^{(r)}(p))\, |\Lambda^{(m_{1}, \ldots, m_{n})}(\varepsilon, p)\rangle &= 
		h^{(r)}_{i}\,
		|\Lambda^{(m_{1}, \ldots, m_{n})}(\varepsilon, p)\rangle \,, \qquad i = 1, \ldots, p \,, 
		\label{eigenvalueproblem}
	\end{align}
	as follows from (\ref{commutativity}) and (\ref{QGsym}). 
	
	The transfer matrix eigenvalues $\Lambda^{(m_{1}, \ldots, 
		m_{n})}(u,\varepsilon, p)$ are given by 
	\be
	\Lambda^{(m_1,...,m_n)}(u, \varepsilon, p)
	=&\phi(u, \varepsilon, p)\, 
	\lambda^{(m_1,...,m_n)}(u,\varepsilon, p) \,, 
	\label{eigenvalue1}
	\ee
	with
	\begin{align}
		\lambda^{(m_1,...,m_n)}(u,\varepsilon, p)=&	
		A(u)\, z_0(u,\varepsilon)\, 
		y_0(u,\varepsilon, p)\, 
		c(u)^{2N}+\tilde{A}(u)\, \tilde{z}_0(u, \varepsilon)\, \tilde{y}_0(u,\varepsilon, p)\, \tilde{c}(u)^{2N}\nonumber\\
		&\hspace{-0.3in} +\Big\{\sum_{l=1}^{n-1}\left[
		z_l(u, \varepsilon)\, y_l(u,\varepsilon,p)\, B_l(u)
		+\tilde{z}_l(u, \varepsilon)\, 
		\tilde{y}_l(u,\varepsilon, p)\, \tilde{B}_l(u)\right]  \nonumber\\
		&\hspace{-0.3in} + z_n(u, \varepsilon)\, 
		y_n(u, \varepsilon, p)\, B_n(u)
		+\tilde{z}_n(u, \varepsilon)\, \tilde{y}_n(u, \varepsilon, p)\, 
		\hat{B}_n(u)\Big\}\, b(u)^{2N}\,.
		\label{eigenvalue2}
	\end{align}
	\noindent
	The overall factor $\phi(u, \varepsilon, p)$ is given by\footnote{For $\varepsilon=0$, the $\phi$ and $y_l$ here are mapped to the ``old'' ones in \cite{Nepomechie:2018nvl} in the following way
		\begin{equation}
			\phi=\phi^{(\text{old})} Z\,, \quad y_{l}=y_l^{(\text{old})}/Z \,, \quad 
			Z =  \frac{\cosh^2(u-n \eta)}{\cosh(u-(n-2p)\eta)\cosh(u-(n+2p)\eta)} \,, \nonumber
		\end{equation}
		so that their products are the same
		$\phi\, y_{l} = \phi^{(\text{old})}\, y_l^{(\text{old})}$.}
	\begin{equation}
		\phi(u,\varepsilon, p)=\frac{\cosh^2(u-n\eta+\frac{\ri \pi \varepsilon}{2})}{\cosh(u+(n-2p)\eta+\frac{\ri \pi \varepsilon}{2})\cosh(u-(3n-2p)\eta+\frac{\ri \pi \varepsilon}{2})} \,.
		\label{phi}
	\end{equation}
	The tilde denotes crossing $\tilde{A}(u)=A(-u-\rho)$, etc.
	The functions $A(u)$ and $B_l(u)$ are given by
	\begin{align}
		& A(u)=\frac{Q^{[1]}(u+\eta)}{Q^{[1]}(u-\eta)}\frac{Q^{[1]}(u+\eta+\ri\pi)}{Q^{[1]}(u-\eta+\ri\pi)}\,, \label{Au}\\
		& B_l(u)=\frac{Q^{[l]}(u-(l+2)\eta)}{Q^{[l]}(u-l\eta)}\frac{Q^{[l]}(u-(l+2)\eta+\ri\pi)}{Q^{[l]}(u-l\eta+\ri\pi)}\nonumber\\
		& \hspace{1.2cm}\times 
		\frac{Q^{[l+1]}(u-(l-1)\eta)}{Q^{[l+1]}(u-(l+1)\eta)}\frac{Q^{[l+1]}(u-(l-1)\eta+\ri\pi)}{Q^{[l+1]}(u-(l+1)\eta+\ri\pi)},\hspace{1cm} l=1,...,n-1\,,\nonumber\\
		& 
		B_n(u)=\frac{Q^{[n]}(u-(n+2)\eta)}{Q^{[n]}(u-n\eta)}\frac{Q^{[n]}(u-(n-2)\eta+\ri\pi)}{Q^{[n]}(u-n\eta+\ri\pi)} \,, \nonumber\\
		& \hat{B}_n(u)= B_{n}(u + \ri \pi) \,, 
		\label{B Dn(2)}
	\end{align}
	\noindent
	where $Q^{[l]}(u)$ are given by
	\begin{align}
		Q^{[l]}(u) &= \prod_{j=1}^{m_l}\sinh\left(\tfrac{1}{2}(u-u_j^{[l]})\right)\sinh\left(\tfrac{1}{2}(u+u_j^{[l]})\right)\,,
		\qquad  Q^{[l]}(-u) = Q^{[l]}(u) \,, \qquad l = 1, \ldots, n-1 \,, 
		\label{Ql} \\
		Q^{[n]}(u) &=  
		\prod_{j=1}^{m_n}\sinh\left(\tfrac{1}{2}(u-u_j^{[n]})\right)\sinh\left(\tfrac{1}{2}(u+u_j^{[n]} 
		+\ri \pi \varepsilon)\right)\,,
		\qquad  Q^{[n]}(-u + \ri\pi \varepsilon) = Q^{[n]}(u) \,.
		\label{Qln}
	\end{align}
	Note the appearance of $\varepsilon$ in the highest-level Q-function \eqref{Qln}, which is a key difference with respect to the case $\varepsilon=0$ \cite{Nepomechie:2018nvl}.
	Another (related) difference with respect to the  $\varepsilon=0$ case is the presence in \eqref{eigenvalue2} of 
	$\hat{B}_n$ instead of $\tilde{B}_n$.
	The zeros $u_j^{[l]}$ of $Q^{[l]}(u)$  (and their number $m_{l}$) are still to be determined. 
	Corresponding results for the closed chain with periodic BCs are given in \cite{Reshetikhin:1987}.
	
	The functions $c(u)$ and $b(u)$ are given by
	\begin{align}
		c(u) &= 4 \sinh(u-2\eta)\, \sinh(u-2 n \eta) \,, \non\\
		b(u) &= 4 \sinh(u)\, \sinh(u-2 n \eta) \,.
		\label{cb}
	\end{align}
	The functions $z_{l}(u, \varepsilon)$ are given by
	\begin{equation}
		z_l(u,\varepsilon)=\begin{cases}
			\frac{2\sinh(2u)\sinh(2u-4n\eta)}{\sinh(2u-2n \eta)}\frac{\sinh(u-(n+1)\eta-\frac{\ri \pi \varepsilon}{2})\cosh(u-(n-1)\eta+\frac{\ri \pi \varepsilon}{2})}{\sinh(2u-2l\eta)\sinh(2u-2(l+1)\eta)}, & 0\le l \le n-1\\
			z_{n-1}(u,\varepsilon)\frac{\sinh(u-(n-1)\eta+\frac{\ri \pi \varepsilon}{2})}{\sinh(u-(n+1)\eta+\frac{\ri \pi \varepsilon}{2})}, & l=n \,,
		\end{cases}
	\end{equation}
	and the functions $y_{l}(u, \varepsilon, p)$ are given by
	\begin{equation}
		y_l(u, \varepsilon, p)=\begin{cases}
			\left( \frac{\cosh(u+(n-2p)\eta+\frac{\ri \pi \varepsilon}{2})}{\cosh(u-n\eta+\frac{\ri \pi \varepsilon}{2})}\right)^2, & 0\le l\le p-1\\
			1, & p\le l\le n
		\end{cases} \,.
		\label{yl}
	\end{equation}
	
	The duality property (\ref{duality}) of the transfer matrix  implies that the corresponding eigenvalues satisfy
	\be
	\Lambda^{(m_1,...,m_n)}(u, \varepsilon, p)= 
	f(u,\varepsilon, p)\, 
	\Lambda^{(m_1,...,m_n)}(u,\varepsilon, n-p) 
	\,.
	\ee
	It follows from (\ref{eigenvalue1}) and (\ref{phi}) that 
	\be
	\lambda^{(m_1,...,m_n)}(u, \varepsilon, p)= \lambda^{(m_1,...,m_n)}(u, \varepsilon, n-p) \,.
	\ee
	
	\subsubsection{Bethe equations}
	
	The Bethe equations can be determined, as usual, from 
	the requirement that the expression for the transfer-matrix 
	eigenvalues (\ref{eigenvalue2}) have vanishing residues at the poles.
	In this way, we find that the Bethe equations are given for $n=1$ with $p=0, 1$ by 
	\begin{equation}
		\left[\frac{\sinh(u_k^{[1]}+\eta)}{\sinh(u_k^{[1]}-\eta)}\right]^{2N}
		=\frac{Q_k^{[1]}\left(u_k^{[1]}+2\eta\right)}{Q_k^{[1]}\left(u_k^{[1]}-2\eta\right)}\,, 
		\quad k = 1, \ldots, m_{1} \,; \label{BEn1}
	\end{equation}
	\noindent
	and for $n>1$ with $p=0, \ldots, n$ by
	\begin{align}
		\left[\frac{\sinh(u_k^{[1]}+\eta)}{\sinh(u_k^{[1]}-\eta)}\right]^{2N}\Phi_{1,\varepsilon,p,n}(u_k^{[1]})&=\frac{Q_k^{[1]}\left(u_k^{[1]}+2\eta\right)}{Q_k^{[1]}\left(u_k^{[1]}-2\eta\right)}\frac{Q_k^{[1]}\left(u_k^{[1]}+2\eta+\ri\pi\right)}{Q_k^{[1]}\left(u_k^{[1]}-2\eta+\ri\pi\right)}\nonumber\\
		&\quad \times 
		\frac{Q^{[2]}\left(u_k^{[1]}-\eta\right)}{Q^{[2]}\left(u_k^{[1]}+\eta\right)}\frac{Q^{[2]}\left(u_k^{[1]}-\eta+\ri\pi\right)}{Q^{[2]}\left(u_k^{[1]}+\eta+\ri\pi\right)}\,, \nonumber\\
		&\quad k=1,\ldots, m_1\,,
		\label{BAl1}
	\end{align}
	\noindent
	\begin{align}
		\Phi_{l,\varepsilon,p,n}(u_k^{[l]})& =\frac{Q^{[l-1]}\left(u_k^{[l]}-\eta\right)}{Q^{[l-1]}\left(u_k^{[l]}+\eta\right)}\frac{Q^{[l-1]}\left(u_k^{[l]}-\eta+\ri\pi\right)}{Q^{[l-1]}
			\left(u_k^{[l]}+\eta+\ri\pi\right)}\nonumber\\
		&\quad \times \frac{Q_k^{[l]}\left(u_k^{[l]}+2\eta\right)}{Q_k^{[l]}\left(u_k^{[l]}-2\eta\right)}\frac{Q_k^{[l]}\left(u_k^{[l]}+2\eta+\ri\pi\right)}{Q_k^{[l]}\left(u_k^{[l]}-2\eta+\ri\pi\right)}\nonumber\\
		&\quad \times 
		\frac{Q^{[l+1]}\left(u_k^{[l]}-\eta\right)}{Q^{[l+1]}\left(u_k^{[l]}+\eta\right)}\frac{Q^{[l+1]}\left(u_k^{[l]}-\eta+\ri\pi\right)}{Q^{[l+1]}\left(u_k^{[l]}+\eta+\ri\pi\right)}\,, \nonumber\\
		&\quad k=1, \ldots, m_l\,, \quad l=2,\ldots, n-1\,,
		\label{BAl}
	\end{align}
	\noindent
	\begin{align}
		\Phi_{n,\varepsilon,p,n}(u_k^{[n]})& 
		=\frac{Q^{[n-1]}\left(u_k^{[n]}-\eta\right)}{Q^{[n-1]}\left(u_k^{[n]}+\eta\right)}\frac{Q^{[n-1]}\left(u_k^{[n]}-\eta+\ri\pi\right)}{Q^{[n-1]}
			\left(u_k^{[n]}+\eta+\ri\pi\right)}\frac{Q_k^{[n]}\left(u_k^{[n]}+2\eta\right)}{Q_k^{[n]}\left(u_k^{[n]}-2\eta\right)}\,, \nonumber\\
		&\quad k=1, \ldots, m_n \,,
		\label{BAln}
	\end{align}
	where $Q^{[l]}(u)$ is given by (\ref{Ql})-(\ref{Qln}), 
	and $Q_k^{[l]}(u)$ is defined by
	\begin{align}
		Q_k^{[l]}(u) &=\prod_{j=1, j \ne 
			k}^{m_l}\sinh\left(\tfrac{1}{2}(u-u_j^{[l]})\right)\sinh\left(\tfrac{1}{2}(u+u_j^{[l]})\right) \,, \qquad l = 1, \ldots, n-1\,, \non\\
		Q_k^{[n]}(u) &=\prod_{j=1, j \ne k}^{m_n}
		\sinh\left(\tfrac{1}{2}(u-u_j^{[n]})\right)
		\sinh\left(\tfrac{1}{2}(u+u_j^{[n]} +\ri \pi \varepsilon)\right) \,.
		\label{Q jk}
	\end{align}
	The factor $\Phi_{l,\varepsilon, p,n}(u)$ is defined by
	\begin{equation}
		\Phi_{l,\varepsilon,p,n}(u)=\left[
		\frac{\cosh\left(u-\delta_{l,p}(n-p)\eta
			+\frac{\ri \pi}{2} \varepsilon \right)}
		{\cosh\left(u+\delta_{l,p}(n-p)\eta
			+\frac{\ri \pi}{2} \varepsilon\right)}\right]^2 \,.
		\label{phi2}
	\end{equation}
	Note that $\Phi_{l,\varepsilon,p,n}(u)$ is different from 1 only if $l=p$.
	
	\subsubsection{Eigenvalues of the Cartan generators}
	
	We assume that the Bethe states 
	$|\Lambda^{(m_{1}, \ldots\,, m_{n})}(\varepsilon,p)\rangle$ are highest-weight states of both the ``left'' and ``right'' algebras \footnote{Compared with \cite{Nepomechie:2018nvl}, here we 
		use a different definition of the ``right'' generators \eqref{eq:rightgenerators}, hence the assumption that the Bethe states are highest (instead of lowest) weights of the ``right'' algebra. A further consequence is the change of sign in $h_i^{(r)}$ in \eqref{CartanCardinalities2} 
		in comparison with Eq. (4.1) in \cite{Nepomechie:2018nvl}, which ensures $h_i^{(r)} \ge 0$.}
	\begin{align}
		\Delta_{N}(E^{+ (\ell)}_{i}(p))\, |\Lambda^{(m_{1}, \ldots, m_{n})}(\varepsilon, p)\rangle &=0 \,, 
		\qquad i = 1, \ldots, n-p\,, \nonumber \\
		\Delta_{N}(E^{+ (r)}_{i}(p))\, |\Lambda^{(m_{1}, \ldots, m_{n})}(\varepsilon, p)\rangle &=0 \,, 
		\qquad i = 1, \ldots, p\,.
		\label{highestweight}
	\end{align}
	The eigenvalues of the Cartan generators $h^{(\ell)}_{i}$ and $h^{(r)}_{i}$ 
	(recall Eq. (\ref{eigenvalueproblem})) are given by 
	\begin{subequations}
		\label{CartanCardinalities}
		\begin{align}
			h_i^{(\ell)} &=  m_{p+i-1}-m_{p+i} \,, &  i &=1,...,n-p \,, \label{CartanCardinalities1} \\
			h_i^{(r)} &=  m_{i-1}-m_i \,,  & i &=1,...,p \,,
			\label{CartanCardinalities2}
		\end{align}
	\end{subequations}
	where $m_0=N$.
	
	The ``left'' and ``right'' Dynkin labels of the Bethe states are 
	given in terms of the eigenvalues of the Cartan generators by
	\begin{align}
		a_i^{(\ell)} &=  h^{(\ell)}_{i} - h^{(\ell)}_{i+1}\,, &\qquad i=1,...,n-p-1 \,,	\non \\
		a_{n-p}^{(\ell)} &=  2h^{(\ell)}_{n-p}\,,
		\label{Dynkinl0}
	\end{align}
	and 
	\begin{align}
		a_i^{(r)} &=  h^{(r)}_{i} - h^{(r)}_{i+1} \,, &\qquad i=1,...,p-1 \,,	\non \\
		a_{p}^{(r)} &=  2 h^{(r)}_{p} \,,
		\label{Dynkinr0}
	\end{align}
	respectively. 
	The Dynkin labels of the Bethe states are therefore given in terms of 
	$m$'s by 
	\begin{align}
		a_i^{(\ell)} &=m_{p+i-1}-2m_{p+i}+m_{p+i+1}\,, &\qquad i=1,...,n-p-1 \,,	\non \\
		a_{n-p}^{(\ell)} &= 2m_{n-1} - 2 m_{n} \,,
		\label{Dynkinl}
	\end{align}
	and 
	\begin{align}
		a_i^{(r)} &=m_{i-1}-2m_{i}+m_{i+1}\,, &\qquad i=1,...,p-1 \,,	\non \\
		a_{p}^{(r)} &= 2m_{p-1} - 2 m_{p} \,.
		\label{Dynkinr}
	\end{align}
	
	Since the Dynkin labels of an irrep determine its
	dimension, these formulas help determine the degeneracies of the transfer-matrix eigenvalues.
	As discussed in \cite{Nepomechie:2018wzp}, the degeneracies can be larger than expected from the QG symmetry, 
	due to additional discrete symmetries, such as self-duality, bonus symmetry, etc.
	
	\subsubsection{Completeness}
	We have numerically verified the completeness of this Bethe ansatz for small values of $n$ and $N$ (for $\varepsilon = 0, 1$ and all $p=0, \ldots, n$) along the lines in \cite{Nepomechie:2017hgw}.
	
	\section{A sketch of the known results for the $D^{(2)}_2$ model}\label{sec:D22}
	
	Unlike the higher-rank models, the scaling limit of the $D^{(2)}_{2}$ model with various boundary conditions is well understood by now. Following the observation \cite{Frahm:2012eb} that the spectrum of the periodic model is identical to that of the so-called staggered six-vertex (or afm Potts) model, it has been shown that the $D^{(2)}_2$ R-matrix as well as the K-matrices leading to the quantum group invariant models with open boundary conditions for $\varepsilon=0,1$ \cite{Martins:2000xie,Nepomechie:2018wzp} can be factorized, i.e.\ expressed in terms of products of these objects for the six-vertex (or $A^{(1)}_1$) model after a similarity transformation \cite{Robertson:2020eri, Nepomechie:2020onv}.\footnote{In fact, the factorization of K-matrices can be generalized to even more general open BCs \cite{Frahm:2022gtk, Li:2022clv}.}  This factorization carries over to the transfer matrices with the corresponding boundary conditions.
	
	The interest in the critical properties of the staggered six-vertex model was sparked by the remarkable work of Ikhlef, Jacobsen and Saleur \cite{Ikhlef:2008zz}, who discovered that the spectrum of scaling dimensions becomes dense in the thermodynamic limit:
	on a finite lattice, a class of $\mathcal{O}(N)$ states is obtained by gradually changing Bethe roots starting from the configuration of the ground state.\footnote{The precise mechanism depends on the model, see e.g.\ \cite{EsFS05,FrMa11,VeJS14,VeJS16a,FrHM19}. For the staggered six-vertex model, the towers are obtained by increasing the disbalance between the numbers of Bethe roots on the real line and with $\Im m(u_k)=\pi$. For the $D^{(2)}_3$ model with boundary conditions $(\varepsilon,p)=(1,1)$, the mechanism is sketched in Table~\ref{tab:towerroots} and Figure~\ref{fig:bc11_towerroots} below.} The energies of these states are separated by gaps $\Delta E\propto 1/(N\log^2(N))$ and extend beyond the low-energy regime.
	In a series of further studies \cite{Ikhlef:2011ay, CaIk13, Frahm:2013cma, BazKotKovLuk2019, BazKotKovLuk2021}  the advanced analytical techniques available for the six-vertex model have been applied to establish that for anisotropies $\eta=\ri\gamma$ in the critical regime $\gamma\in (0,\tfrac{\pi}{2})$ the low-energy physics is described by the 2D black hole sigma model, a conformal field theory with non-compact target originally introduced by Witten \cite{Witten:1991yr}.
	In addition to the continuum of exponents, this model also possesses a series of discrete states \cite{Dijkgraaf:1991ba, Hanany:2002ev, RiSc04}. In the lattice model, states from the continuum and discrete ones transform into each other under a twist \cite{VeJS14,FrHo17,BazKotKovLuk2021}. This relates to the truncation of the discrete series of levels in the CFT by unitarity.
	
	The finite-size spectrum of the $D^{(2)}_2$ model has also been analyzed for anti-diagonal closed \cite{FrGe2024} and the two quantum-group invariant open BCs. Studies of the latter, originally motivated by the possibility to identify non-compact branes for the 2D black hole CFTs, led to the following observation: the lattice model with boundary conditions $\varepsilon=0$ is in the same universality class as the afm Potts model with central charge
	\begin{subequations}
		\label{eq:PottsCFT}
		\begin{equation}
			c=2-6\frac{\gamma}{\pi}\,,
		\end{equation}
		and has a purely discrete spectrum of conformal weights \cite{Robertson:2020eri}
		\begin{equation}
			\label{eq:paraF}
			h_{\mathcal{S}}=\frac\gamma\pi\,\mathcal{S}\left(\mathcal{S}+1\right)\,,
		\end{equation}
	\end{subequations}
	labeled by the $U_q(B_1)$ spin $\mathcal{S}$ of primaries.\footnote{For integer $k=\pi/\gamma$ (\ref{eq:PottsCFT}) are the central charge and a subset of the conformal spectrum of the $Z_{k-2}$ parafermion CFT \cite{Saleur91}.} 
	Only for $\varepsilon=1$ the continuous component of the spectrum is preserved. In the scaling limit the spectrum of effective conformal dimensions has been found to be \cite{Robertson:2020imc,FrGe22,FrGK24}
	\begin{subequations}
		\label{eq:D22weights}
		\begin{align}
			\label{eq:D22cont}
			X_{\text{eff}} = -\frac{c}{24}+h_{\mathcal{S},s}=-\frac1{12} + \frac\gamma{4\pi}\left(2\mathcal{S}+1-\frac\pi{\gamma}\right)^2 + \frac\gamma{\pi-2\gamma}\,s^2\,,
		\end{align}
		where the real parameter $s$ labels the continuous components of the spectrum.  Note that (\ref{eq:D22cont}) implies that the ground state of the model has a nonzero spin depending on the anisotropy $\gamma$, and therefore the symmetry of the model is spontaneously broken.  Under variation of $\gamma$ the lowest states in the continuum undergo the transmutation into discrete ones. The corresponding conformal weights are again given by (\ref{eq:D22cont}) but now $s$ takes the discrete imaginary values \cite{FrGe22,FrGK24}
		\begin{align}
			\label{eq:D22disc}
			s = \pm \ri \left(\mathcal{S}+1+a-\frac\pi{2\gamma}\right)\,,
			\quad a=0,1,2,\dots<\frac\pi{2\gamma}-(\mathcal{S}+1)\,.
		\end{align}
	\end{subequations}
	The characterization of the scaling limit is completed by an explicit formula for the density of states characterizing the continuous spectrum computed in \cite{FrGK24}.  The resulting expression for the partition function, however, does not seem to correspond to known results in the literature on branes in the 2D black hole CFTs.

	\section{The $D^{(2)}_3$ model}\label{sec:D23}
	In the rest of the manuscript we focus on the case where 
	\begin{equation}
		n=2\,,
	\end{equation}
	and therefore the local Hilbert space at each site has dimension $2n+2=6$.
	
	\subsection{Boundary conditions of interest and their symmetries}\label{subsec:symmetries}
	
	Recall that $\varepsilon$ and $p$ can take the values
	$\varepsilon=0,1$ and $p=0,..., n$; and duality  \eqref{duality} relates $p$ and $n-p$. For $n=2$, we can therefore restrict our attention to the following four cases:
	\begin{align}
		(\varepsilon,p)= (0,0) &: \quad U_q(B_2)  &&+ \, Z_2  \text{ symmetry} \,; \nonumber \\[0.1 cm]
		(\varepsilon,p)= (0,1) &: \quad U_q(B_1)\otimes  U_{q}(B_1) &&+ Z_2  \quad +\text{self-duality symmetry} \,; \nonumber \\[0.1 cm]
		(\varepsilon,p)= (1,0) &: \quad U_q(B_2) &&+ Z_2  \text{ symmetry}\,;&&  \nonumber \\[0.1 cm]
		(\varepsilon,p)= (1,1) &: \quad U_q(B_1)\otimes  U_{q}(B_1) &&+ Z_2  \quad + \text{self-duality + bonus symmetry} \,.
		\label{cases}
	\end{align}
	For each case, the symmetries are listed, as follows from Section \ref{subsec:transf}.
	All four cases have the $Z_2$ symmetry \eqref{Z2symmetry},
	as for the closed chain \cite{Frahm:2023lpe}. We conjectured in \cite{Frahm:2023lpe} that this symmetry shifts all type-2 Bethe roots by $\ri \pi$, and a similar result could also hold for the open chain.
	
	For the closed chain, we found an additional 
	$Z_2$ symmetry if the twist angles satisfy $\phi_1 + \phi_2=0$, see (2.25) in  \cite{Frahm:2023lpe}. We can now understand that symmetry as a special case of the \emph{self-duality} symmetry \eqref{selfdual} that appears for the open chain with $p=1$ (and both values of $\varepsilon$). Indeed, for the open chain with $p=1$, the self-duality relates the two copies of $U_q(B_1)$;
	for the closed chain, the $U_q(B_1)$'s are broken to $U(1)$'s, which are related by the $Z_2$ symmetry (see (2.24) in \cite{Frahm:2023lpe}), leading to the constraint $\phi_1 + \phi_2=0$.  In other words, for $p=1$ (and both values of $\varepsilon$), 
	the self-duality symmetry plays the role of the $Z_2$ symmetry given by (2.25) in  \cite{Frahm:2023lpe}. (For $p=0$, there is no such symmetry, since duality relates $p=0$ to $p=2$.)
	
	For the case $(\varepsilon,p)= (1,1)$, in addition to self-duality, there is also bonus symmetry. The self-duality and bonus symmetries presumably form a larger finite (discrete) group, which remains to be identified.

	\subsection{The Hamiltonian of the $D^{(2)}_3$ model}
	
	In order to investigate the ground state and low-lying excitations, it is necessary to first unambiguously define the Hamiltonian.
	To this end, we observe that the expression for 
	the eigenvalues $\Lambda(u,\varepsilon, p)$ of the transfer matrix $t(u,\varepsilon, p)$ \eqref{transfer} in terms of Bethe roots
	is given by 
	\begin{align}
		& \Lambda(u,\varepsilon, p) =	
		A(u)\, q_0(u,\varepsilon, p)\, c(u)^{2N}
		+\tilde{A}(u)\, \tilde{q}_0(u,\varepsilon, p)\, \tilde{c}(u)^{2N}\nonumber\\
		& + \left[ q_1(u,\varepsilon, p)\,  B_1(u)
		+\tilde{q}_1(u,\varepsilon, p)\, \tilde{B}_1(u)
		+  q_2(u,\varepsilon, p)\, B_2(u) 
		+ \tilde{q}_2(u,\varepsilon, p)\, \hat{B}_2(u) \right]\, b(u)^{2N}\,,
		\label{Lambda}
	\end{align}
	see Eqs. \eqref{eigenvalue1} and \eqref{eigenvalue2},
	where the functions $q_l(u,\varepsilon, p)$ are defined by
	\begin{equation}
		q_l(u,\varepsilon, p) = \phi(u,\varepsilon, p)\, z_l(u,\varepsilon)\, y_l(u,\varepsilon, p) \,.
	\end{equation}
	We wish to define the Hamiltonian in such a way that its eigenvalues in terms of the Bethe roots are given by 
	\begin{equation}
		E=\sum^{m_1}_{k=1} \varepsilon_0(u^{[1]}_k)=-\sum^{m_1}_{k=1} \frac{2\sinh^2(2\eta)}{\cosh(2u^{[1]}_k)-\cosh(2\eta)}\,,
		\label{energy}
	\end{equation}
	which is the same as for the twisted $D^{(2)}_2$ and $D^{(2)}_3$ models. We note that 
	\begin{equation}
		E = \frac{1}{2}\sinh(2\eta)
		\frac{A'(0)}{A(0)} \,, \qquad b(0)=\tilde{c}(0)=0 \,,
		\label{energy2}
	\end{equation}
	see Eqs. \eqref{Au}, \eqref{Ql}, \eqref{cb}.
	We now proceed to construct the Hamiltonians for the four cases \eqref{cases}.
	
	\subsubsection{$(\varepsilon, p) \ne (1,1)$}
	
	For the three cases $(\varepsilon, p) \ne (1,1)$, the function $q_0(u,\varepsilon, p)$ is nonzero for $u=0$. We see that 
	\begin{equation}
		\Lambda(0,\varepsilon,p) = q_0(0,\varepsilon, p)\, c(0)^{2N}\,, 
		\qquad (\varepsilon, p) \ne (1,1) \,,
	\end{equation}
	and the energy \eqref{energy2} is given in terms of the first derivative of the transfer-matrix eigenvalue \eqref{Lambda} at $u=0$
	\begin{equation}
		E = \frac{1}{2}\sinh(2\eta) \left[ 
		\frac{\Lambda'(0,\varepsilon,p)}{\Lambda(0,\varepsilon,p)}
		- \frac{q'_0(0,\varepsilon,p)}{q_0(0,\varepsilon,p)}
		- \frac{2N c'(0)}{c(0)} \right]\,, \qquad (\varepsilon, p) \ne (1,1) \,.
		\label{ELambda}
	\end{equation}
	We therefore define the Hamiltonian $\mathcal{H}$ in terms of the transfer matrix by
	\begin{equation}
		\mathcal{H} = \frac{\sinh(2\eta) }{2\Lambda(0,\varepsilon,p)}
		t'(0,\varepsilon,p)
		-  \frac{1}{2}\sinh(2\eta) \left[ 
		\frac{q'_0(0,\varepsilon,p)}{q_0(0,\varepsilon,p)}
		+ \frac{2N c'(0)}{c(0)} \right]\id \,, \qquad (\varepsilon, p) \ne (1,1) \,.
	\end{equation}
	We observe, following Sklyanin \cite{Sklyanin:1988yz}, that
	\begin{align}
		t'(0,\varepsilon,p) &= \xi(0)^{2N}\Big\{
		\frac{2\left( \tr_a K^L_a(0,\varepsilon,p) \right)}{\xi(0)}  \mathrm{h} 
		+  \left( \tr_a K^L_a(0,\varepsilon,p) \right)  K^{R'}_1(0,\varepsilon,p) \nonumber\\
		&  + \frac{2}{\xi(0)} \tr_a K^L_a(0,\varepsilon,p) h_{N,a}   
		+   \tr_a K^{L'}_a (0,\varepsilon,p) \Big\} \,,
	\end{align}
	where 
	\begin{equation}
		\mathrm{h}  = \sum_{k=1}^{N-1} h_{k,k+1}\,, \qquad h_{i,j} = {\cal P}_{i,j}\, R'_{i,j}(0) \,,
		\label{hdefinitions}
	\end{equation}
	and
	\begin{equation}
		\xi(u)=4\sinh(u+2\eta)\sinh(u+4\eta).
		\label{eq:xi}
	\end{equation}
	We conclude that the Hamiltonian is given by
	\begin{align}
		\mathcal{H} &= \frac{\sinh(2\eta) }{2q_0(0,\varepsilon, p)}
		\Big\{  \frac{2\left( \tr_a K^L_a(0,\varepsilon,p) \right)}{c(0)}  \mathrm{h} + \left( \tr_a K^L_a(0,\varepsilon,p) \right)  K^{R'}_1(0,\varepsilon,p) 
		+  \frac{2}{c(0)} \tr_a K^L_a(0,\varepsilon,p) h_{N,a} \nonumber\\
		& + \left[  \tr_a K^{L'}_a (0,\varepsilon,p) 
		- q'_0(0,\varepsilon,p)
		-\frac{2N c'(0)\, q_0(0,\varepsilon,p)}{c(0)} \right]\id \Big\}\,, \qquad (\varepsilon, p) \ne (1,1) \,.
		\label{Hamiltonian1}
	\end{align}

	In the isotropic limit $\eta \rightarrow 0$, the expression for the Hamiltonian \eqref{Hamiltonian1} becomes quite simple
	\begin{equation}
		\mathcal{H}\big\vert_{\eta \rightarrow 0} = \begin{cases}
			-\frac{1}{2}G +\frac{3}{2}(N-1) \id  & (\varepsilon, p) = (0,0), (0,1) \\[0.1in]
			b_1  -\frac{1}{2}G + b_N +\frac{3}{2}(N-1) \id  & (\varepsilon, p) = (1,0)
		\end{cases} \,,
	\end{equation}
	with 
	\begin{equation}
		G = \sum_{k=1}^{N-1} g_{k,k+1}\,, 
		\label{Gdef}
	\end{equation}
	where the 2-site Hamiltonian is given by
	\begin{equation}
		g = \id + 2\mathcal{P} - \mathcal{K}\,, 
		\qquad \mathcal{P} = \sum_{a,b=1}^6 e_{ab}\otimes e_{ba} \,,
		\qquad \mathcal{K} = \sum_{a,b=1}^6 e_{ab}\otimes e_{7-a,7-b} \,, 
		\label{g2site}
	\end{equation}
	and the 1-site boundary term is given by
	\begin{equation}
		b = \frac{1}{2}\left(e_{33} + e_{44} - e_{34} - e_{43} \right) \,.
	\end{equation}
	The ferromagnetic state $|0\rangle=((1,0,0,0,0,0)^t)^{\otimes N}$ is a ground state, with  zero energy, of the isotropic Hamiltonian
	\begin{equation}
		\mathcal{H}\big\vert_{\eta \rightarrow 0} |0\rangle = 0 \,.
	\end{equation}
	
	\subsubsection{$(\varepsilon, p) = (1,1)$}
	
	For the case $(\varepsilon, p)= (1,1)$, the function $q_0(u,\varepsilon, p)$ is zero for $u=0$, and therefore the transfer-matrix-eigenvalue vanishes $\Lambda(0,\varepsilon=p=1)=0$. Indeed, the transfer matrix is zero for $u=0$, since 
	\begin{equation}
		\tr_a K^L_a(0,\varepsilon=p=1)=0 \,.
	\end{equation}
	Moreover, 
	\begin{equation}
		\Lambda'(0,\varepsilon=p=1) = A(0)\, q'_0(0,\varepsilon=p=1)\, c(0)^{2N} \,.
	\end{equation}
	Hence, as is the case for $D^{(2)}_2$ with $\varepsilon=1$ \cite{Robertson:2020imc, Nepomechie:2020onv},
	the expression for the energy \eqref{energy2} is now given in terms of the {\it second} derivative of the transfer-matrix eigenvalue at $u=0$
	\begin{equation}
		E = \frac{1}{2}\sinh(2\eta) \left[ 
		\frac{\Lambda''(0,\varepsilon=p=1)}{2 \Lambda'(0,\varepsilon=p=1)}
		- \frac{q''_0(0,\varepsilon=p=1)}{2q'_0(0,\varepsilon=p=1)}
		- \frac{2N c'(0)}{c(0)} \right] \,,
	\end{equation}
	c.f. \eqref{ELambda}.
	We therefore define the Hamiltonian in terms of the transfer matrix by
	\begin{equation}
		\mathcal{H} = \frac{1}{4}\sinh(2\eta)\, t'(0)^{-1}\,
		t''(0)  -  \frac{1}{2}\sinh(2\eta) \left[ 
		\frac{q''_0(0)}{2q'_0(0)}
		+ \frac{2N c'(0)}{c(0)} \right]\id  \,,
	\end{equation} 
	where it is now understood that $\varepsilon=p=1$.
	After a long computation, we obtain
	\begin{align}
		t''(0) &= \xi(0)^{2N}\Big[
		\frac{2}{\xi(0)^2}\big\{\big\{ \mathrm{h}\,, \mathrm{k}^{(0)} \big\} \,, K^R_1(0) \big\}
		- \frac{8 \sinh(2\eta)}{\xi(0)}\big\{ \mathrm{h}\,, K^R_1(0) \big\}\nonumber\\
		&+ 8\cosh(2\eta) K^R_1(0) - 8 \sinh(2\eta)  K^{R'}_1(0) 
		+ \frac{4}{\xi(0)} \mathrm{k}^{(2)}\, K^R_1(0) \nonumber\\
		&+ \frac{2}{\xi(0)} \mathrm{k}^{(1)}\, K^R_1(0) 
		+ \frac{2}{\xi(0)^2} \mathrm{k}^{(3)}\, K^R_1(0) 
		+ \frac{4}{\xi(0)} \mathrm{k}^{(0)}\, K^{R'}_1(0) \Big] \,,
	\end{align}
	where, in addition to \eqref{hdefinitions}, we have defined
	\begin{align}
		\mathrm{k}^{(0)} &= \tr_a K^L_a(0) h_{N,a} \,, \quad
		\mathrm{k}^{(1)} = \tr_a K^L_a(0) h'_{N,a} \,, 
		\quad h'_{N,a} = {\cal P}_{N,a} R_{N,a}''(0)\,, \nonumber\\
		\mathrm{k}^{(2)} &= \tr_a K^{L'}_a(0) h_{N,a} \,, \quad
		\mathrm{k}^{(3)} = \tr_a K^L_a(0) h_{N,a}^2 \,, 
	\end{align}
	and $\{ \,, \}$ denotes the anti-commutator. Moreover, we note the identities
	\begin{align}
		t'(0)^{-1} &=  \frac{1}{\xi(0)^{2N}}  \frac{\tanh(4\eta)}{2-4 \cosh(2\eta)} K^R_1(0)\, K^R_N(0) \,, \non \\
		K^R(0)^{-1} &= K^R(0) = 
		\tiny\setlength{\arraycolsep}{1pt}\left( \begin{array}{c c;{2pt/2pt}c c;{2pt/2pt}c c}
			1 & 0 &  &  &  & \\ 
			0 &-1 &  &  &  & \\ \hdashline[2pt/2pt]
			& & 0 & -1 &  & \\
			& & -1 & 0 &  &\\ \hdashline[2pt/2pt]
			& &  &  & -1 & 0\\ 
			& &  &  & 0 & 1
		\end{array}\right)\,.
	\end{align}
	We conclude that the Hamiltonian for $\varepsilon=p=1$ is given by
	\begin{align}
		\mathcal{H} &= \frac{\sinh(2\eta)}{4}   
		\frac{\tanh(4\eta)}{2-4 \cosh(2\eta)} K^R_1(0)\, K^R_N(0)
		\Big[
		\frac{2}{\xi(0)^2}\big\{\big\{ \mathrm{h}\,, \mathrm{k}^{(0)} \big\} \,, K^R_1(0) \big\}
		- \frac{8 \sinh(2\eta)}{\xi(0)}\big\{ \mathrm{h}\,, K^R_1(0) \big\}\nonumber\\
		&+ 8\cosh(2\eta) K^R_1(0) - 8 \sinh(2\eta)  K^{R'}_1(0) 
		+ \frac{4}{\xi(0)} \mathrm{k}^{(2)}\, K^R_1(0) \nonumber\\
		&+ \frac{2}{\xi(0)} \mathrm{k}^{(1)}\, K^R_1(0) 
		+ \frac{2}{\xi(0)^2} \mathrm{k}^{(3)}\, K^R_1(0) 
		+ \frac{4}{\xi(0)} \mathrm{k}^{(0)}\, K^{R'}_1(0) \Big]
		-  \frac{1}{2}\sinh(2\eta) \left[ 
		\frac{q''_0(0)}{2q'_0(0)}
		+ \frac{2N c'(0)}{c(0)} \right]\id \,.
		\label{Hamiltonian2}
	\end{align}
	
	In the isotropic limit $\eta \rightarrow 0$, the expression for the Hamiltonian \eqref{Hamiltonian2} simplifies considerably
	\begin{align}
		\mathcal{H}\big\vert_{\eta\rightarrow 0}  &= -\frac{1}{8}\left(G 
		+ K^R_1(0)\, G\, K^R_1(0) + K^R_N(0)\, G\, K^R_N(0) 
		+ K^R_1(0) K^R_N(0)\, G\, K^R_1(0) K^R_N(0) \non \right) \non \\
		&+\frac{3}{2}(N-1) \id \,,
	\end{align}
	where $G$ is given by \eqref{Gdef}.
	The ground states of the isotropic Hamiltonian, with zero energy
	\begin{equation}
		\mathcal{H}\big\vert_{\eta \rightarrow 0} |0\rangle = 0 \,,
		\label{ferro}
	\end{equation}
	span the ferromagnetic multiplet (including the pseudo vacuum in the sector $(h^{(\ell)},h^{(r)})=(0,N)$) with degeneracy $\text{dim}\left(\left([0]\otimes [N]\right) \oplus \left([N]\otimes [0]\right)\right)=2(2N+1)$. Here and below, we use the notation $[\mathcal{S}]$ to denote a spin-$\mathcal{S}$ representation of $B_1$, which has dimension $2\mathcal{S}+1$.

	\subsection{Bethe equations for $D^{(2)}_3$}
	In the rest of the manuscript, we restrict ourselves to the parametric regime 
	\begin{align}
		\eta=\ri \gamma \qquad\text{where}\qquad   \gamma\in (0,\tfrac{\pi}{4}),\,\label{mfsoleme}
	\end{align}
	which has been investigated for the periodic boundary conditions \cite{Frahm:2023lpe}. Below the explicit form  of the Bethe equations of the open $D^{(2)}_3$ chain in the parametrization \eqref{mfsoleme} are given for the different integrable boundary conditions parameterized by $\varepsilon,\,p=0,\,1$, 
	see Eqs. \eqref{BAl1}-\eqref{phi2} and \eqref{CartanCardinalities}. 
	Changes compared to the $(\varepsilon,p)=(0,0)$ case are displayed in red. Recall that all BCs have the same energy functional (\ref{energy}) in terms of the Bethe roots \footnote{For the BC $(\varepsilon,p)=(1,1)$, the possible presence of singular strings leads to a modification of this formula, see \eqref{energy11} below.}:
	\begin{equation}
		E = -\sum_{j=1}^{m_1} \frac{2\sinh^2(2\ri\gamma)}{\cosh(2u^{[1]}_j)-\cos(2\gamma)}\,.
		\label{energyagain}
	\end{equation}
	
	%===================================================================
	\subsubsection{$(\varepsilon,p)=(0,0)$:}
	\begin{equation}
		\label{bae00}
		\begin{aligned}
			\left(\frac{\sinh(u^{[1]}_j+\ri\gamma)}{\sinh(u^{[1]}_j-i\gamma)}\right)^{2N}
			=& \prod_{k\neq j}^{m_1}
			\left(\frac{\sinh(u^{[1]}_j-u^{[1]}_k+2\ri\gamma)}{\sinh(u^{[1]}_j-u^{[1]}_k-2\ri\gamma)}\, 
			\frac{\sinh(u^{[1]}_j+u^{[1]}_k+2\ri\gamma)}{\sinh(u^{[1]}_j+u^{[1]}_k-2\ri\gamma)}\right)\times\\
			&\times \prod_{k=1}^{m_2} \left(\frac{\sinh(u^{[1]}_j-u^{[2]}_k-\ri\gamma)}{\sinh(u^{[1]}_j-u^{[2]}_k+\ri\gamma)}\,
			\frac{\sinh(u^{[1]}_j+u^{[2]}_k-\ri\gamma)}{\sinh(u^{[1]}_j+u^{[2]}_k+\ri\gamma)}\right)\,,\qquad\quad\,\, j=1\dots m_1\\
			1=&\prod_{k=1}^{m_1} \left(\frac{\sinh(u^{[2]}_j-u^{[1]}_k-\ri\gamma)}{\sinh(u^{[2]}_j-u^{[1]}_k+\ri\gamma)}\,
			\frac{\sinh(u^{[2]}_j+u^{[1]}_k-\ri\gamma)}{\sinh(u^{[2]}_j+u^{[1]}_k+\ri\gamma)}\right)\times\\
			&\times \prod_{k\neq j}^{m_2}
			\left(\frac{\sinh\frac12(u^{[2]}_j-u^{[2]}_k+2\ri\gamma)}{\sinh\frac12(u^{[2]}_j-u^{[2]}_k-2\ri\gamma)}\, 
			\frac{\sinh\frac12(u^{[2]}_j+u^{[2]}_k+2\ri\gamma)}{\sinh\frac12(u^{[2]}_j+u^{[2]}_k-2\ri\gamma)}\right)\,,
			\quad j=1\dots m_2
		\end{aligned}
	\end{equation}
	with $m_1=N-h_1$, $m_2=N-h_1-h_2$. (To lighten the notation, we denote the eigenvalues of the Cartan generators $h^{(\ell)}_j$ as $h_j$.)
	
	%===================================================================
	\subsubsection{$(\varepsilon,p)=(0,1)$:}
	\begin{equation}
		\label{bae01}
		\begin{aligned}
			\left(\frac{\sinh(u^{[1]}_j+\ri\gamma)}{\sinh(u^{[1]}_j-\ri\gamma)}\right)^{2N}
			=& {\color{red}\left(\frac{\cosh(u^{[1]}_j+\ri\gamma)}{\cosh(u^{[1]}_j-\ri\gamma)}\right)^2}
			\prod_{k\neq j}^{m_1}
			\left(\frac{\sinh(u^{[1]}_j-u^{[1]}_k+2\ri\gamma)}{\sinh(u^{[1]}_j-u^{[1]}_k-2\ri\gamma)}\, 
			\frac{\sinh(u^{[1]}_j+u^{[1]}_k+2\ri\gamma)}{\sinh(u^{[1]}_j+u^{[1]}_k-2\ri\gamma)}\right)\times\\
			&\times \prod_{k=1}^{m_2} \left(\frac{\sinh(u^{[1]}_j-u^{[2]}_k-\ri\gamma)}{\sinh(u^{[1]}_j-u^{[2]}_k+\ri\gamma)}\,
			\frac{\sinh(u^{[1]}_j+u^{[2]}_k-\ri\gamma)}{\sinh(u^{[1]}_j+u^{[2]}_k+\ri\gamma)}\right)\,,\qquad\quad\,\, j=1\dots m_1\\
			1=&\prod_{k=1}^{m_1} \left(\frac{\sinh(u^{[2]}_j-u^{[1]}_k-\ri\gamma)}{\sinh(u^{[2]}_j-u^{[1]}_k+\ri\gamma)}\,
			\frac{\sinh(u^{[2]}_j+u^{[1]}_k-\ri\gamma)}{\sinh(u^{[2]}_j+u^{[1]}_k+\ri\gamma)}\right)\times\\
			&\times \prod_{k\neq j}^{m_2}
			\left(\frac{\sinh\frac12(u^{[2]}_j-u^{[2]}_k+2\ri\gamma)}{\sinh\frac12(u^{[2]}_j-u^{[2]}_k-2\ri\gamma)}\, 
			\frac{\sinh\frac12(u^{[2]}_j+u^{[2]}_k+2\ri\gamma)}{\sinh\frac12(u^{[2]}_j+u^{[2]}_k-2\ri\gamma)}\right)\,,
			\quad j=1\dots m_2
		\end{aligned}
	\end{equation}
	where $m_1=N-h^{(r)}$, $m_2=N-h^{(r)}-h^{(\ell)}$. (To lighten the notation, we drop the subscripts of the eigenvalues of the Cartan generators, and therefore denote $h^{(\ell)}_1$ and $h^{(r)}_1$
	as $h^{(\ell)}$ and $h^{(r)}$, respectively.)
	
	%===================================================================
	\subsubsection{$(\varepsilon,p)=(1,0)$:}
	\begin{equation}
		\label{bae10}
		\begin{aligned}
			\left(\frac{\sinh(u^{[1]}_j+\ri\gamma)}{\sinh(u^{[1]}_j-\ri\gamma)}\right)^{2N}
			=& \prod_{k\neq j}^{m_1}
			\left(\frac{\sinh(u^{[1]}_j-u^{[1]}_k+2\ri\gamma)}{\sinh(u^{[1]}_j-u^{[1]}_k-2\ri\gamma)}\, 
			\frac{\sinh(u^{[1]}_j+u^{[1]}_k+2\ri\gamma)}{\sinh(u^{[1]}_j+u^{[1]}_k-2\ri\gamma)}\right)\times\\
			&\times \prod_{k=1}^{m_2} \left(\frac{\sinh(u^{[1]}_j-u^{[2]}_k-\ri\gamma)}{\sinh(u^{[1]}_j-u^{[2]}_k+\ri\gamma)}\,
			\frac{\sinh(u^{[1]}_j+u^{[2]}_k-\ri\gamma)}{\sinh(u^{[1]}_j+u^{[2]}_k+\ri\gamma)}\right)\,,\qquad\quad\,\, j=1\dots m_1\\
			1=&\prod_{k=1}^{m_1} \left(\frac{\sinh(u^{[2]}_j-u^{[1]}_k-\ri\gamma)}{\sinh(u^{[2]}_j-u^{[1]}_k+\ri\gamma)}\,
			\frac{\sinh(u^{[2]}_j+u^{[1]}_k-\ri\gamma)}{\sinh(u^{[2]}_j+u^{[1]}_k+\ri\gamma)}\right)\times\\
			&\times \prod_{k\neq j}^{m_2}
			\left(\frac{\sinh\frac12(u^{[2]}_j-u^{[2]}_k+2\ri\gamma)}{\sinh\frac12(u^{[2]}_j-u^{[2]}_k-2\ri\gamma)}\, 
			\frac{{\color{red}\cosh}\frac12(u^{[2]}_j+u^{[2]}_k+2\ri\gamma)}{{\color{red}\cosh}\frac12(u^{[2]}_j+u^{[2]}_k-2\ri\gamma)}\right)\,,
			\quad j=1\dots m_2
		\end{aligned}
	\end{equation}
	with $m_1=N-h_1$, $m_2=N-h_1-h_2$. 
	
	%============================================================
	\subsubsection{$(\varepsilon,p)=(1,1)$:}
	\label{sssec:bae11}
	\begin{equation}
		\label{bae11}
		\begin{aligned}
			\left(\frac{\sinh(u^{[1]}_j+\ri\gamma)}{\sinh(u^{[1]}_j-\ri\gamma)}\right)^{2N\color{red}-2}
			=& \prod_{k\neq j}^{m_1}
			\left(\frac{\sinh(u^{[1]}_j-u^{[1]}_k+2\ri\gamma)}{\sinh(u^{[1]}_j-u^{[1]}_k-2\ri\gamma)}\, 
			\frac{\sinh(u^{[1]}_j+u^{[1]}_k+2\ri\gamma)}{\sinh(u^{[1]}_j+u^{[1]}_k-2\ri\gamma)}\right)\times\\
			&\times \prod_{k=1}^{m_2} \left(\frac{\sinh(u^{[1]}_j-u^{[2]}_k-\ri\gamma)}{\sinh(u^{[1]}_j-u^{[2]}_k+\ri\gamma)}\,
			\frac{\sinh(u^{[1]}_j+u^{[2]}_k-\ri\gamma)}{\sinh(u^{[1]}_j+u^{[2]}_k+\ri\gamma)}\right)\,,\qquad\quad\,\,  j=1\dots m_1\\
			1=&\prod_{k=1}^{m_1} \left(\frac{\sinh(u^{[2]}_j-u^{[1]}_k-\ri\gamma)}{\sinh(u^{[2]}_j-u^{[1]}_k+\ri\gamma)}\,
			\frac{\sinh(u^{[2]}_j+u^{[1]}_k-\ri\gamma)}{\sinh(u^{[2]}_j+u^{[1]}_k+\ri\gamma)}\right)\times\\
			&\times \prod_{k\neq j}^{m_2}
			\left(\frac{\sinh\frac12(u^{[2]}_j-u^{[2]}_k+2\ri\gamma)}{\sinh\frac12(u^{[2]}_j-u^{[2]}_k-2\ri\gamma)}\,            \frac{{\color{red}\cosh}\frac12(u^{[2]}_j+u^{[2]}_k+2\ri\gamma)}{{\color{red}\cosh}\frac12(u^{[2]}_j+u^{[2]}_k-2\ri\gamma)}\right)\,,
			\quad j=1\dots m_2
		\end{aligned}
	\end{equation}
	where $m_1=N-h^{(r)}$, $m_2=N-h^{(r)}-h^{(\ell)}$.
	
	An important feature of the BC $(\varepsilon,p)=(1,1)$ 
	is the existence of eigenstates whose Bethe roots include
	one or more so-called \emph{singular} roots 
	of type-1 and/or type-2.\footnote{This includes the ground state for $\pi/6<\gamma<\pi/4$ for small $N$: the ground state of the system with up to $N=6$ sites is a doublet, $2\cdot\left([0]\otimes[0]\right)$ in terms of the $U_q(B_1)$-spins, and its root configuration contains one pair of singular roots on level-1 ($n^{[1]}_{\text{sing}}=2$) together with a pair of  complex conjugate level-2 roots on the imaginary axis.  Around $\gamma=\pi/5$ we observe a level crossing between $N=6$ and $8$ after which the ground state becomes tenfold degenerate ($\left([0]\otimes [2]\right) \oplus \left([2]\otimes [0]\right)$) with a root configuration containing strings (\ref{d23-strings}) only (no singular roots), as discussed in Sect.~\ref{ssec:fsa11} below.} 
	The type-1 singular roots are exact 1-strings given by
	\begin{subequations}
		\label{sing12}
		\begin{align}
			u^{[1]}_j =  \ri\,\gamma\, \quad \text{for~} j=1, \ldots,  n^{[1]}_{\text{sing}}\,,
			\label{sing}
		\end{align}
		and the type-2 singular roots are arranged  in exact 3-strings given by
		\begin{align}
			\BR{2}{j,k} = \ri\,k\,\gamma\,, \quad k=0,\pm2\,
			\quad \text{for~} j=1, \ldots, n^{[2]}_{\text{sing}}\,,
			\label{sing2}
		\end{align}
	\end{subequations}
	where $n^{[1]}_{\text{sing}}$ and $n^{[2]}_{\text{sing}}$ denote the number of type-1 and type-2 singular strings, respectively.\footnote{We do not rule out the possible appearance of longer singular strings, which however we have not encountered in our investigation of low-lying states.}
	Note that despite the close similarity of these Bethe equations to those for $(\varepsilon,p)=(1,0)$ in \eqref{bae10}, we have not found any instances of Bethe roots with singular strings \eqref{sing12} there.
	
	Singular Bethe roots give rise to apparent zeros or poles in the Bethe equations, and poles in the energy formula \eqref{energy}, but nevertheless lead to analytic transfer matrix eigenvalues.
	These singular roots must be treated separately from the others such that the Bethe eqs.\ (\ref{bae11}) become
	\begin{equation}
		\label{bae11sing}
		\begin{aligned}
			\left(\frac{\sinh(u^{[1]}_j+\ri\gamma)}{\sinh(u^{[1]}_j-\ri\gamma)}\right)^{2N\color{red}-2-n^{[1]}_{\text{sing}}}
			&= {\color{red}\left(\frac{\sinh(u^{[1]}_j+3\ri\gamma)}{\sinh(u^{[1]}_j-3\ri\gamma)}\right)^{n^{[1]}_{\text{sing}}-2n^{[2]}_{\text{sing}}}
			}\times\\
			&\times \prod_{{\color{red}k>n^{[1]}_{\text{sing}}},{k\neq j}}^{m_1}
			\left(\frac{\sinh(u^{[1]}_j-u^{[1]}_k+2\ri\gamma)}{\sinh(u^{[1]}_j-u^{[1]}_k-2\ri\gamma)}\, 
			\frac{\sinh(u^{[1]}_j+u^{[1]}_k+2\ri\gamma)}{\sinh(u^{[1]}_j+u^{[1]}_k-2\ri\gamma)}\right)\times\\
			&\times \prod^{m_2}_{\color{red}k>3n^{[1]}_{\text{sing}}} \left(\frac{\sinh(u^{[1]}_j-u^{[2]}_k-\ri\gamma)}{\sinh(u^{[1]}_j-u^{[1]}_k+\ri\gamma)}\,
			\frac{\sinh(u^{[1]}_j+u^{[2]}_k-\ri\gamma)}{\sinh(u^{[1]}_j+u^{[2]}_k+\ri\gamma)}\right)\,,\\
			&\hspace{6cm}j=n^{[1]}_{\text{sing}}+1, \ldots, m_1\\
			1&={\color{red}\left(\frac{\sinh(u^{[2]}_j-2\ri\gamma)}{\sinh(u^{[2]}_j+2\ri\gamma)}\right)^{n^{[1]}_\text{sing}-n^{[2]}_\text{sing}}
				\left(\frac{\sinh(u^{[2]}_j+4\ri\gamma)}
				{\sinh(u^{[2]}_j-4\ri\gamma)}\right)^{n^{[2]}_\text{sing}}
			}\times\\
			&\times \prod_{{\color{red}k>n^{[1]}_{\text{sing}}}}^{m_1} \left(\frac{\sinh(u^{[2]}_j-u^{[1]}_k-\ri\gamma)}{\sinh(u^{[2]}_j-u^{[1]}_k+\ri\gamma)}\,
			\frac{\sinh(u^{[2]}_j+u^{[1]}_k-\ri\gamma)}{\sinh(u^{[2]}_j+u^{[1]}_k+\ri\gamma)}\right)\times\\
			&\times \prod_{{\color{red}k>3n^{[2]}_{\text{sing}}}, k\neq j}^{m_2}
			\left(\frac{\sinh\frac12(u^{[2]}_j-u^{[2]}_k+2\ri\gamma)}{\sinh\frac12(u^{[2]}_j-u^{[2]}_k-2\ri\gamma)}\, 
			\frac{{\color{red}\cosh}\frac12(u^{[2]}_j+u^{[2]}_k+2\ri\gamma)}{{\color{red}\cosh}\frac12(u^{[2]}_j+u^{[2]}_k-2\ri\gamma)}\right)\,,\\
			&\hspace{6cm} j=3n^{[2]}_{\text{sing}}+1, \ldots, m_2
		\end{aligned}
	\end{equation}
	The contribution of the singular roots to the energy (\ref{energy}) can be given explicitly (for $N>2$):\footnote{This result can be derived using the energy expression \eqref{energy2} and the expression for $A(u)$ (see Eqs.  \eqref{Au} and \eqref{Ql}), and assuming that there are $n^{[1]}_{\text{sing}}$ singular level-$1$ Bethe roots \eqref{sing}. This formula does not hold for the special case with $N=2$ and $n^{[1]}_{\text{sing}}=2$, in which case $\Lambda''(0)$ has an additional contribution involving $B_1(u)$.}
	\begin{equation}
		E_{(\varepsilon,p)=(1,1)} = {\color{red}n^{[1]}_{\text{sing}}\cos(2\gamma)} - \sum_{j=n^{[1]}_{\text{sing}}+1}^{m_1} \frac{2\sinh^2(2\ri\gamma)}{\cosh(2u^{[1]}_j)-\cos(2\gamma)}\,.
		\label{energy11}
	\end{equation}

	\subsection{The thermodynamic limit} 
	The periodic model in the parameter range \eqref{mfsoleme}
	is critical \cite{Frahm:2023lpe}. We expect that the property of being at criticality is mainly a bulk property and so is unchanged by the change of BCs. 
	For a one-dimensional open critical lattice model it is expected that the spectrum of low-energy excitations can be described within the framework of a two-dimensional boundary conformal field theory. The following prediction for large-$N$ asymptotics of the energy is expected to hold \cite{BlCN86, Cardy:1986gw} from conformal field theory
	\begin{align}
		E\asymp N e_\infty+f_\infty +\frac{\pi v_F}{N} \left(-\frac{c}{24}+\Delta+{\tt d}\right)\label{fmekmemfke}\,.
	\end{align}
	Here, $e_\infty$ is the bulk energy density, $f_\infty$ is the surface energy, while $v_F$ is the Fermi velocity. These all are model-dependent quantities. The universal conformal field theory content is contained in the $N^{-1}$ correction in \eqref{fmekmemfke}. We denote by $c$ the central charge, by $\Delta$ the dimensions of the conformal primary fields, while ${\tt d}$ denotes the level of the descendant.  For the further analysis, it is useful to define the effective scaling dimension (or effective conformal weight)
	\begin{align}\label{fmskmdskrr}
		X_{\rm eff}=-\frac{c}{24}+\Delta+{\tt d}\,.
	\end{align}
	
	The methodology used in the rest of the paper is to use the Bethe ansatz to numerically extract the energy $E$ of a given Bethe state for larger and larger system sizes\footnote{We kindly refer the reader to section 3 of \cite{Frahm:2023lpe} where it is explained how a given Bethe state is extended to higher system sizes.} and then use the following finite-size estimate
	\begin{align}\label{flslsnnvkvk}
		\frac{N }{\pi v_F}(E-Ne_\infty-f_\infty) \,
	\end{align}
	to extract the allowed values of \eqref{fmskmdskrr}. We remark here that the Hamiltonian is not Hermitian. Nevertheless, we find numerically that for all low-energy states (i.e. states for which $E-E_{GS}\sim \frac{1}{N}$) we have considered, the imaginary part of the energies decays to zero faster than $1/N$, yielding that the conformal data is real as usual, leading to a sensible low-energy theory.
	
	The first step to proceed via the above is to calculate the non-universal quantities, namely, the bulk energy density $e_\infty$, the surface energy density $f_\infty$ and the Fermi velocity $v_F$. This is done by
	using the root-density formalism for the open model as sketched in the following. 
	
	For all boundary conditions $(\varepsilon,p)$, a small-system analysis by exact diagonalization yields that the root configuration for the ground state consists of the following 2-strings on both levels, similar to the periodic model \cite{Frahm:2023lpe}:
	\begin{equation}
		\label{d23-strings}
		u^{[1]}_j = x_j \pm \ri\left( \frac\pi2-\gamma-\delta_j\right)\,,\qquad\quad
		u^{[2]}_j = y_j \pm \frac{\ri\pi}{2}\,.
	\end{equation}
	Here $\delta_j$ are some deviations from the lines with imaginary parts $\pm(\frac\pi2-\gamma)$. We find numerically that these deviations tend to zero as the system size $N$ approaches infinity. Further, we find that the real centers $x_j$ and $y_j$ become dense on the positive real line. The Fourier transform of the bulk density is given by 
	\begin{align}
		\sigma^x(\omega)=\sigma^y(\omega)=\frac{1}{\cosh((2\gamma-\frac{\pi}{2})\omega)}\,.\label{fmvkmcvncvnjs}
	\end{align}
	
	As expected, we find for the bulk quantities $e_\infty$ and velocity $v_F$ the same expression as in the periodic model \cite{Frahm:2023lpe}:
	\begin{align}
		v_F=\frac{\pi  \sin (2 \gamma )}{\pi -4 \gamma },\, \qquad e_\infty=-\frac{1}{2}\int^{\infty}_{-\infty}\text{d}\omega \, \frac{\sin(2\gamma)\sinh(2\gamma\omega)}{\cosh((2\gamma-\frac{\pi}{2})\omega)\sinh(\frac{\omega\pi}{2})} \,.
	\end{align}
	The surface energy $f_\infty$, however,  depends on the BCs imposed. We have that 
	\begin{align}
		\label{finf-allbc}
		f_\infty=-\frac{\sin(2\gamma)}{2}\int^{\infty}_{-\infty}{\rm d}\omega \frac{\sinh(2\gamma \omega)}{\sinh(\frac{\pi\omega}{2})}\, \tau^x(\omega) \,,
	\end{align}
	where $\tau^x(\omega)$ is the boundary contribution to the density of level-1 strings. We give explicit formulae for $f_\infty$ for the different BCs in the following sections. 
	
	We would like to highlight a subtle technical point concerning the root-density approach for the real centers of strings in the open case. In this setting, the scattering phase involving the level-1 roots require a careful treatment. If one inserts the string hypothesis directly, without accounting for deviations, the result will be incorrect. The correct procedure is to first include the strings with small deviations, compute the relevant quantities, and only then take the limit where the deviations vanish. We illustrate this procedure for the case $(\varepsilon,p) = (0, 0)$ in Appendix~\ref{fmkdmngmocmf}; the other cases of BCs follow analogously.

	\subsection{Finite-size analysis} 
	We will not give a general classification of the Bethe root configurations leading to specific conformal weights. This is simply due to the fact that, unfortunately, there is in general no clear structure. We will discuss certain clear excitation mechanisms when possible. In addition, we provide access to some representative examples of the Bethe roots used in this section in the online repository found under \cite{D23.data}. 
	\subsubsection{$(\varepsilon, p)= (0,0)$}
	
	We have that the surface contribution to the root density takes for this particular BCs the following form
	\begin{align}\label{fmdkfdkfdknfjn}
		\tau^x(\omega)= \frac{\big(1+\cosh(\gamma\omega)-\cosh((3\gamma-\frac{\pi}{2})\omega)\big)}{\cosh(\gamma\omega)\cosh((2\gamma-\frac{\pi}{2})\omega)} \,.
	\end{align}
	This yields the surface energy: 
	\begin{align}
		f_\infty&=-\frac{\sin^2(2\gamma)}{\cos(2\gamma)}-\frac{\sin(2\gamma)}{4}\int^{\infty}_{-\infty}\text{d}\omega \, \, \frac{\big(1+\cosh(\gamma\omega)-\cosh((3\gamma-\frac{\pi}{2})\omega)\big)\sinh(2\gamma\omega)}{\cosh(\gamma\omega)\cosh((2\gamma-\frac{\pi}{2})\omega)\sinh(\frac{\pi\omega}{2})} \,.
	\end{align}
	The ground state is a singlet, $h_1=h_2=0$, and the corresponding effective scaling dimension is given by
	\begin{align}
		X_{\rm eff}^{(0)} = -\frac{1}{6} + \frac{10\gamma}{4\pi}-\frac{\gamma }{\pi -2 \gamma }\,.
	\end{align}
	By considering various excited states, we conjecture that the low-energy states give rise to the effective scaling dimensions: 
	\begin{align}\label{fmdkfmdkcndm}
		X_{\rm eff}= -\frac{1}{6} +\frac{\gamma}{\pi}\bigg(h^{}_1+\frac{3}{2}\bigg)^2+\frac{\gamma}{\pi}\bigg(h^{}_2+\frac{1}{2}\bigg)^2-\frac{\gamma }{\pi -2 \gamma }+ {\tt d}\,,
	\end{align}
	where ${\tt d}\in \mathbb{N}$.
	We note that the dependence on the Cartan charges can be rewritten in terms of the quadratic Casimir operator of the symmetry algebra: in the classical limit its eigenvalue on a highest weight state $(h_1,h_2)$ is
	\begin{equation*}
		\mathcal{C}^{B_2}(h_1,h_2)= 2\left( h_1(h_1+3) + h_2(h_2+1)\right)
		=2\left(\left(h_1+\frac32\right)^2 + \left(h_2+\frac12\right)^2\right) - 5\,.
	\end{equation*}
	Examples of the numerical data for the Bethe roots leading to the critical exponents (\ref{fmdkfmdkcndm}) are given in the online repository \cite{D23.data}. 
	
	\subsubsection{$(\varepsilon, p)= (0,1)$} 
	For this particular BCs we have
	\begin{align}
		\tau^x(\omega)=\frac{2 \sinh (\gamma  \omega ) \left(\cosh (\gamma  \omega )-\cosh \left(\frac{1}{2} (\pi -2 \gamma ) \omega \right)+1\right)}{\sinh \left(\frac{1}{2} (8 \gamma -\pi ) \omega \right)+\sinh \left(\frac{\pi  \omega }{2}\right)} \,.
	\end{align}
	This leads to the surface energy: 
	\begin{align}
		f_\infty=-\frac{\sin(2\gamma)}{2}\int^{\infty}_{-\infty}{\rm d}\omega \frac{\sinh(2\gamma \omega)}{\sinh(\frac{\pi\omega}{2})}\, \frac{2 \sinh (\gamma  \omega ) \left(\cosh (\gamma  \omega )-\cosh \left(\frac{1}{2} (\pi -2 \gamma ) \omega \right)+1\right)}{\sinh \left(\frac{1}{2} (8 \gamma -\pi ) \omega \right)+\sinh \left(\frac{\pi  \omega }{2}\right)} \,.
	\end{align}
	For the ground state, we have $h^{(\ell)}=h^{(r)}=0$ and
	\begin{align}
		\label{eq:ceff01}
		X^{(0)}_{\rm eff}=- \frac{1}{6}+\frac{\gamma}{2\pi}\,.
	\end{align}
	In general, we conjecture that the conformal spectrum can be described by the symmetry algebra, i.e., the left and right $U_q(B_1)$ spins $\mathcal{S}^{(\alpha)}=h^{(\alpha)}$:
	\begin{align}
		\label{eq:Xeff01}
		X_{\rm eff}=- \frac{1}{6}+\frac{\gamma}{4\pi}\bigg(2\mathcal{S}^{(\ell)}+1\bigg)^2+\frac{\gamma}{4\pi}\bigg(2\mathcal{S}^{(r)}+1\bigg)^2+ {\tt d}\,.
	\end{align}
	Comparing this expression to the conformal spectrum of the $D^{(2)}_2$ model $U_q(B_1)$ quantum group symmetry for $\varepsilon=0$ obtained in \cite{Robertson:2020eri} we conjecture that the scaling limit of the $D^{(2)}_3$ model for $(\varepsilon,p)=(0,1)$ BC can be described by two antiferromagnetic Potts models with free boundaries, each contributing (\ref{eq:PottsCFT}) to the central charge and the conformal weights. Further, the degeneracy of the conformal primaries, which we have investigated, is consistent with the product of characters of two independent antiferromagnetic Potts models.
	
	Examples of the numerical data for the Bethe roots are given in the online repository \cite{D23.data}.

	\subsubsection{$(\varepsilon, p)= (1,0)$} 
	For this BC, we find that the surface root density takes the more complicated form
	\be
	\begin{aligned}
		\tau^x(\omega)=&\frac{\text{csch}\left(\frac{\pi  \omega }{2}\right) \sinh \left(\frac{1}{2} (\pi -8 \gamma ) \omega \right)+\text{csch}\left(\frac{\pi  \omega }{4}\right) \sinh \left(\frac{1}{4} (\pi -8 \gamma ) \omega \right)}{\text{csch}\left(\frac{\pi  \omega }{2}\right) \sinh \left(\frac{1}{2} (\pi -8 \gamma ) \omega \right)-1}\\
		&+ \frac{2 \cosh (\gamma  \omega )-2 \coth \left(\frac{\pi  \omega }{2}\right) \sinh (\gamma  \omega )}{\text{csch}\left(\frac{\pi  \omega }{2}\right) \sinh \left(\frac{1}{2} (\pi -8 \gamma ) \omega \right)-1}\,.
	\end{aligned}
	\ee
	such that the corresponding surface root density \eqref{finf-allbc} is given by
	\be
	\begin{aligned}
		f_\infty=&\frac{\sin(2\gamma)}{2}\int^{\infty}_{-\infty}{\rm d}\omega \, \frac{\sinh\left(\frac12(\pi-2\gamma)\omega\right)+2\cosh\left(\frac14(\pi-4\gamma)\omega\right)\cosh(\gamma\omega)\sinh\left(\frac14{(\pi-8\gamma)}\omega\right)}{\cosh\left(\frac12(\pi-4\gamma)\omega\right)\sinh\left(\frac{\omega\pi}{2}\right)}\\
		&+\frac{\sin^2(2\gamma)}{\cos(2\gamma)} \,,
	\end{aligned}
	\ee
	
	By employing formula \eqref{CartanCardinalities1}, we deduce the ground state lies within the sector characterized by $h^{}_1$ in the $\gamma$ interval
	\be
	\frac{\pi}{2(h^{}_1+2)}<\gamma<\frac{\pi}{2(h^{}_1+1)}\,.\label{fmndnfjal}
	\ee
	A numerical analysis confirms these bounds. Hence, the symmetry algebra is spontaneously broken. For the finite-size analysis, we group the low-energy states into two classes A and B. 
	
\begin{figure}[ht] \centerline{\includegraphics[width=0.8\textwidth]{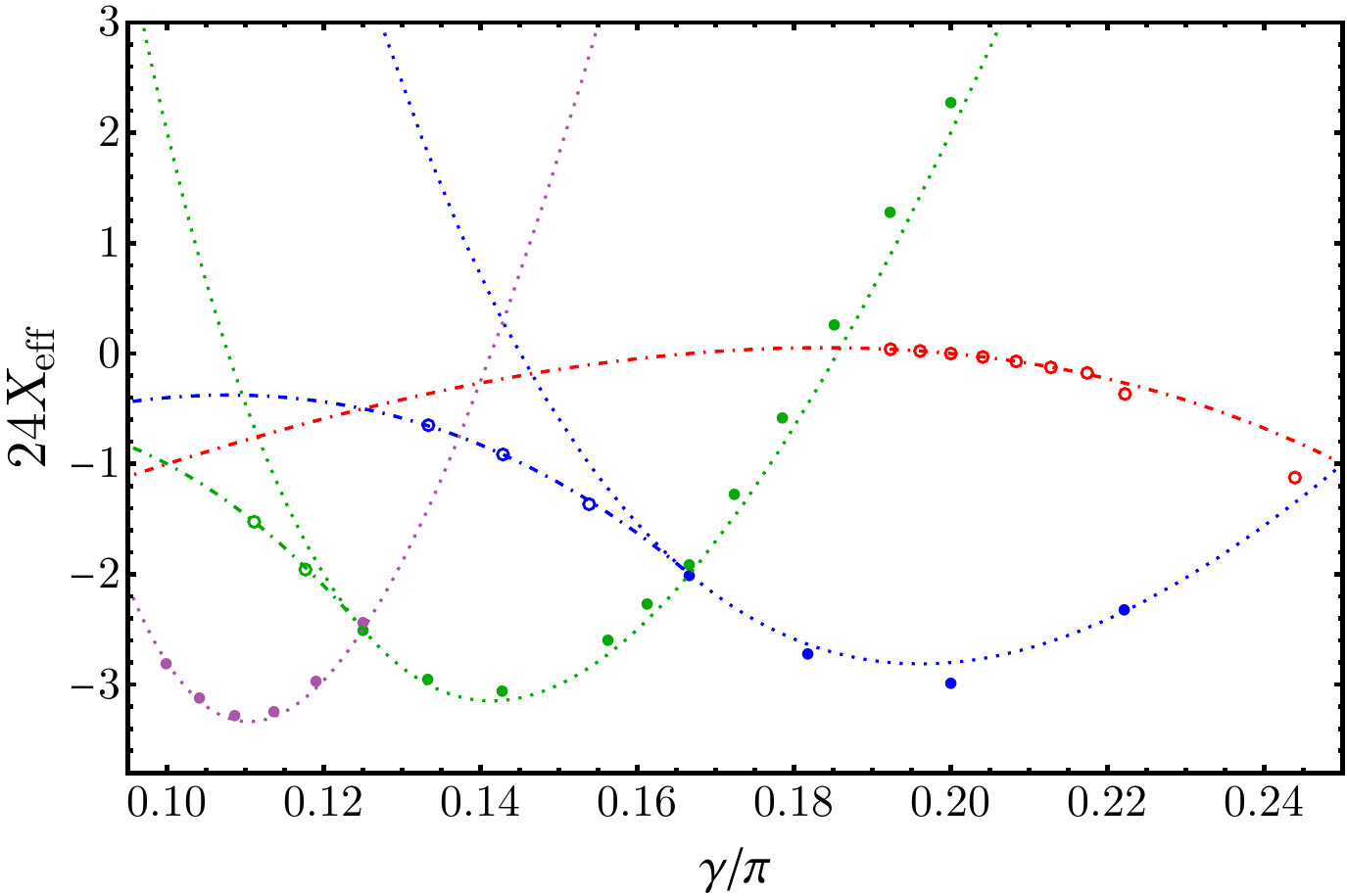}}
		\caption{Effective conformal weight $24 X_{\text{eff}}=\frac{24N}{\pi v_{\rm F}}(E(N)-N e_\infty-f_\infty)$ 
			for states with different $h_1$ of the $D^{(2)}_3$ chain with boundary conditions $(\varepsilon,p)=(1,0)$. The circles display numerical data obtain from the Bethe ansatz.
			The dotted lines display the lowest state in the $(h_1,0)$ continuum yielding logarithmic correction flowing to  \eqref{njdnjsleme} for $h_1=1,2,3$ for blue, green and purple. The dashed states are discrete states \eqref{ndmfndmf2} having power law corrections (red $h_1=0$).   Note that the ground state is never in the $h_1=0$ sector. \label{fmefemkfmek} } 
	\end{figure}

\begin{figure}[ht] \centerline{\includegraphics[width=0.8\textwidth]{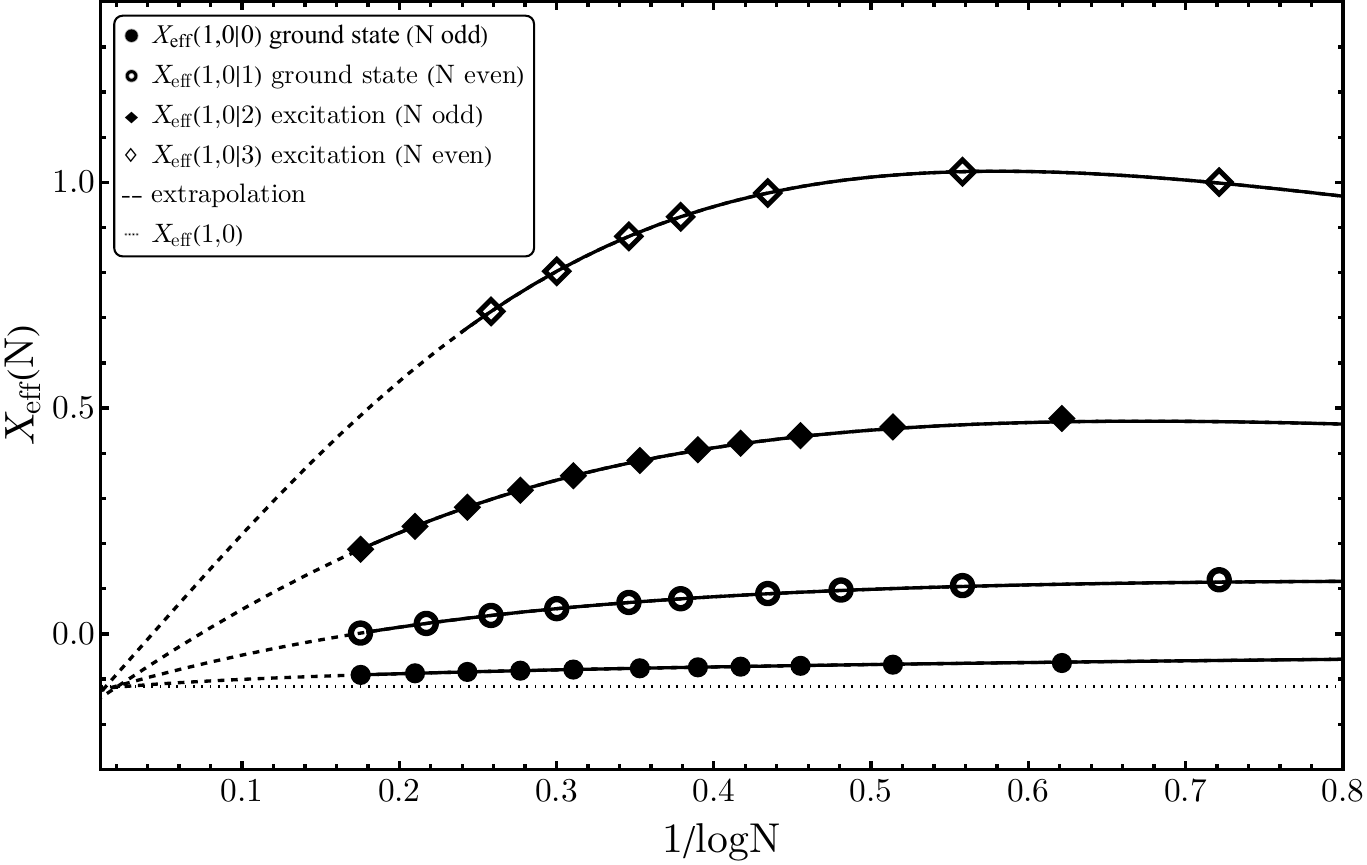}}
		\caption{Effective conformal weight $X_{\text{eff}}=\frac{N}{\pi v_{\rm F}}(E(N)-N e_\infty-f_\infty)$ 
			for excited states  in the sector $h_1=1,h_2=0$ of the $D^{(2)}_3$ chain with boundary conditions $(\varepsilon,p)=(1,0)$. We see strong logarithmic corrections. 
			\label{fhhhhhh} }
	\end{figure}

	\paragraph{Class A:} In a given $h_1$, $h_2$ sector, we find that in the interval \eqref{fmndnfjal} and \eqref{mfsoleme}, the lowest-energetic state consisting just of roots of the type \eqref{d23-strings} possesses the following effective scaling dimension 
	\begin{align}\label{njdnjsleme}
		X_{\rm eff}(h_1,h_2)= -\frac16 + \frac{\gamma}{\pi}\left(h^{}_1 + \frac32-\frac{\pi}{2\gamma} \right)^2+\frac{\gamma}{\pi}\left(h^{}_2 + \frac12 \right)^2\,.
	\end{align}
	Combining \eqref{njdnjsleme} and \eqref{fmndnfjal} yields the effective scaling dimensions of the ground state to be
	\be
	X_{\rm eff}^{(0)}=-\frac16+\frac{\gamma}{4\pi} + \frac{\gamma}{\pi}\left(\text{frac}\left(\frac{\pi}{2\gamma}\right) - \frac12 \right)^2\,,
	\ee
	where $\text{frac}(x)$ denotes the factional part of $x$. Some numerical data is displayed in Fig.~\ref{fmefemkfmek}. 
	Above the states \eqref{njdnjsleme} in the interval \eqref{fmndnfjal} for $h_1=1$ and $h_2=0$, we have numerically estimated the effective scaling dimension of three other states labeled as ${\tt n}=1,2,3$, each one generated by the same excitation mechanism present in the periodic model, i.e. by resolving  four-strings and placing the roots on the lines with fixed imaginary part:
	\begin{align}
		\Im m(u^{[1]}_j)=\frac{\pi}{2}\,,\qquad\quad  \Im m(u^{[2]}_j)=0,\pi\,.
	\end{align}
	For a sketch of the Bethe root configurations, see\footnote{This figure is actually for the BC $(\varepsilon,p)=(1,1)$. However, while the values of the Bethe roots for
		$(\varepsilon,p)=(1,0)$ are different, the qualitative picture is the same.} the Fig.~\ref{fig:bc11_towerroots}.  We have found that these three states possess logarithmic corrections, see Fig.~\ref{fhhhhhh}.
	Based on this analysis and the presence of other states with similar Bethe root configurations for small system-sizes, it is expected that there are an extensive number of such states possessing such logarithmic corrections. Based on our numerical data for ${\tt n}=1,2,3$ we conjecture 
	\begin{align}\label{fmkfmdjfndjfnd}
		X_\text{eff}(h_1,0|{\tt n})= X_{\rm eff}(h_1,0) + {(2\,{\tt n}+1)^2}\,\frac{A(\gamma)}{\log^2(N/N_0)}+\mathcal{O}(N^{-x})\,\quad{\text{for~}{\tt n}=1,2,3,\dots\,.}
	\end{align}
	where ${\tt n}$ is the number of excitations in the continuum tower, and $x$ stands for any remaining power-law decay.  Further, $N_0$ is some non-universal length scale, and $A(\gamma)$ is an amplitude that we do not attempt to calculate here. The amplitude of the logarithmic corrections for the lowest state in the continuum is too small to distinguish it from the power-law behavior in the numerical data. Therefore, we analyze the finite-size differences between the states ${\tt n} = 2$ and ${\tt n} = 0$, and compare them with those between ${\tt n} = 3$ and ${\tt n} = 1$. From this comparison, we conjecture that \eqref{fmkfmdjfndjfnd} also holds for ${\tt n}= 0$. Examples of the numerical data for the Bethe roots are given in the online repository \cite{D23.data}. 
	
	If we fix $h_1$ and $h_2$ and vary the anisotropy $\gamma$ below the lower bound of \eqref{fmndnfjal} and we follow the states \eqref{njdnjsleme} or the higher continuum states, we find transmutation from the continuum states into discrete ones, also seen in Fig.~\ref{fmefemkfmek}. 
	When $\gamma$ is lowered below $\frac{\pi}{2(h_1+2)}$ for fixed $h_1$, the effective scaling dimension of the lowest state in the continuum in the sector $h_1$ changes to
	\begin{align}\label{ndmfndmf2}
		X^{*}_{\rm eff}(h_1,h_2)=X_{\rm eff}(h_1,h_2)- \frac{\gamma}{\pi - 2\gamma}\left(2+h_1 - \frac{\pi}{2\gamma}\right)^2\,.
	\end{align}
	The logarithmic corrections are not present anymore. These states just possess power law corrections to scaling. On the level of the Bethe roots, the transmutation can be seen as a significant change of the Bethe roots:  we see that some of the roots of the continuum state tend to infinity as $\gamma$ approaches the lower bound \eqref{fmndnfjal}. Then the roots come back to a finite real part, but the imaginary part is changed. Also here, we did not manage to find a universal parameterization of these discrete states in terms of the Bethe roots. Hence, we provide examples of the numerical data for the Bethe roots in the online repository \cite{D23.data}. 
	
	\paragraph{Class B: One non-matching state}
	We have identified in addition one single state which possesses no logarithmic corrections but which does not fit in the family \eqref{ndmfndmf2}. It has a conformal dimension of
	\begin{align}
		X^{**}_{\rm eff}(h_1=1,h_2=0)=X_{\rm eff}(h_1=1,h_2=0)-\frac{\gamma}{\pi-2\gamma}\left(2-\frac\pi{2\gamma}\right)^2 +1\,.
	\end{align}

	\subsubsection{$(\varepsilon, p)= (1,1)$} 
	\label{ssec:fsa11}
	Adapting the root-density approach as presented in Appendix~\ref{fmkdmngmocmf} to the present case, the boundary contribution to the density of level-1 strings (\ref{d23-strings}) is found to be
	\begin{align}
		\label{eq:tau11}
		\tau^x(\omega) = -\frac{2 \cosh (\gamma  \omega ) (\cosh (\gamma  \omega )+1)+\tanh \left(\frac{\pi  \omega }{4}\right) \sinh (2 \gamma  \omega )+2 \coth \left(\frac{\pi  \omega }{2}\right) \sinh (\gamma  \omega )}{\text{csch}\left(\frac{\pi  \omega }{2}\right) \sinh \left(\frac{1}{2} (8 \gamma -\pi ) \omega \right)+1}\,,
	\end{align}
	which gives a diverging contribution $\frac12 - (\pi/2\gamma)$ to the total number of corresponding roots as $\gamma\to0$ (cf.\ the $D^{(2)}_2$-chain with quantum-group-invariant boundary conditions \cite{Robertson:2020imc,FrGe22,FrGK24}).  This implies spontaneous broken symmetry of the $D^{(2)}_3$-chain with boundary conditions $(\varepsilon,p)=(1,1)$, as its ground state is realized in the sector $(h^{(\ell)}=0,h^{(r)})$, i.e.\ $m_1=m_2=N-h^{(r)}$, with $h^{(r)}=2,3,4\dots$ for
	\begin{equation}
		\label{eq:intervals11}
		\frac\pi{2h^{(r)}+2}<\gamma < \frac\pi{2h^{(r)}}\,.
	\end{equation}
	Note that in the limit $\gamma\to0$ (or $\gamma<\pi/2N$ for sufficient large finite systems), the ground state is the reference state with $m_1=m_2=0$, see (\ref{ferro}).
	
	With (\ref{eq:tau11}), the surface energy (\ref{finf-allbc}) of the $D^{(2)}_3$ model with boundary conditions $(\varepsilon,p)=(1,1)$ becomes
	\begin{align}
		f_\infty&=\frac{\sin(2\gamma)}{2}\int^{\infty}_{-\infty}{\rm d}\omega\,\frac{2\left(\cosh(\gamma\omega)+1\right)\,\sinh(\frac12(\pi-2\gamma)\omega)+\sinh(2\gamma\omega)}{2\sinh(\frac{\pi}{2}\omega)\,\cosh(\frac12(\pi-4\gamma)\omega)}\,.
	\end{align}
	
	With the predicted scaling of energies (\ref{fmekmemfke}) we can identify the operator content of the boundary CFT describing the scaling limit of the lattice model from its finite-size spectrum.  It turns out that the effective conformal weights can be separated into two contributions corresponding to the two factors in its symmetry algebra $U_q(B_1)\otimes U_q(B_1)$, similar as in (\ref{eq:Xeff01}) for BC $(\varepsilon,p)=(0,1)$.  Based on the characteristics of these contributions, we identify three classes of conformal weights.
	
\begin{figure}[t]
		\centerline{\includegraphics[width=0.8\textwidth]{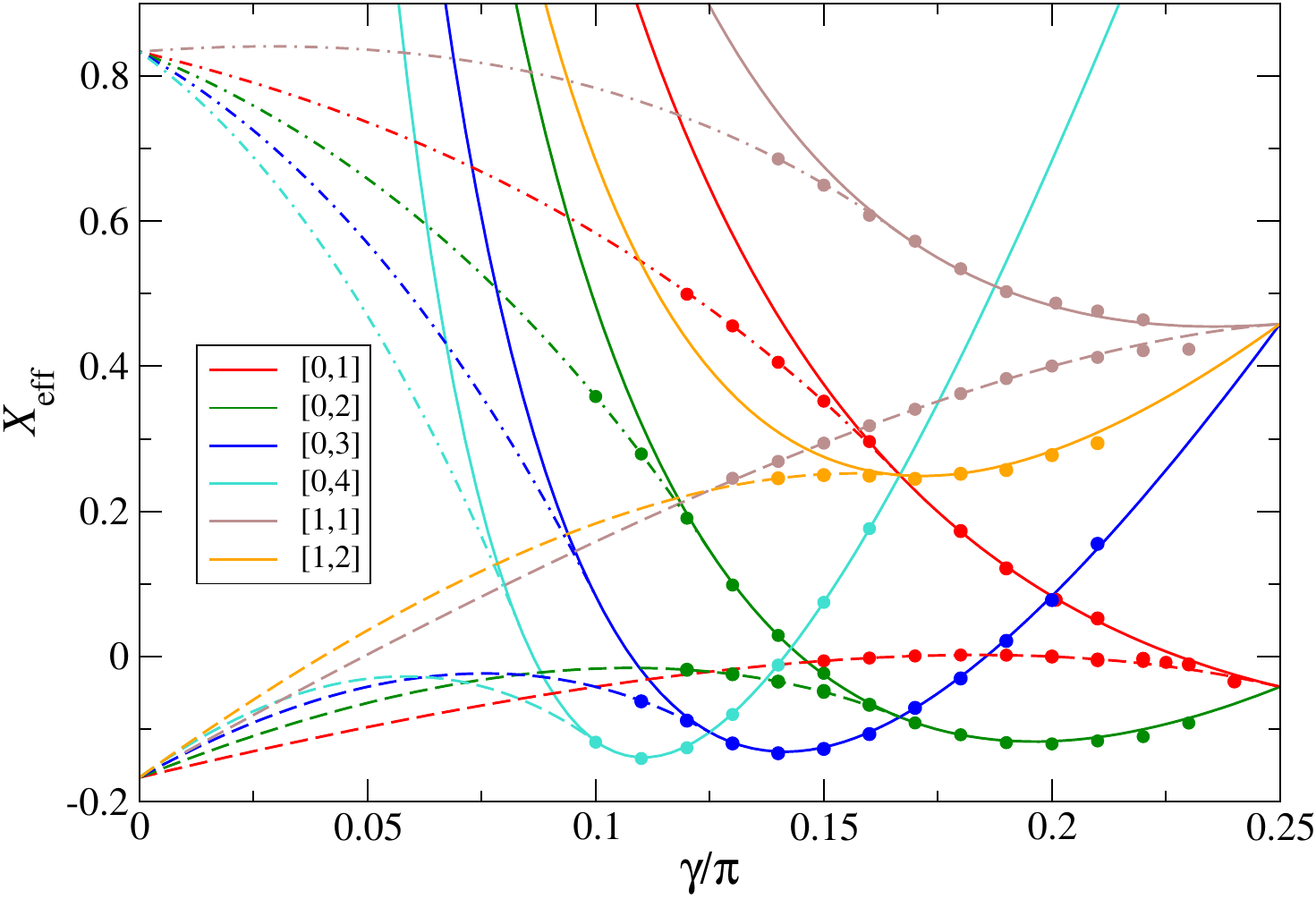}}
		\caption{Effective conformal weights $X_{\text{eff}}$
			for some of the lowest class~A states of the $D^{(2)}_3$ chain with boundary conditions $(\varepsilon,p)=(1,1)$: full lines indicate the lower bounds (\ref{eq:ceff11-k}) the continua in sectors $[\mathcal{S}^{(\ell)},\mathcal{S}^{(r)}]$, dashed and dash-dotted lines are the conjectures $X_\text{eff}^*$ and $X_\text{eff}^{**}$ for the discrete levels emerging from these continua as given in (\ref{eq:xeff11_discrete}), bullets are extrapolations of the numerical finite size data \cite{D23.data}.
			\label{fig:bc11_gs}}
	\end{figure}

	\paragraph{Class A: signatures of a non-compact critical degree of freedom.}
	Based on our numerical solution of the Bethe equations (\ref{bae11}), we conjecture that the lowest state for anisotropies (\ref{eq:intervals11}) in the Cartan sector $(h^{(\ell)},h^{(r)})$ is realized for lattice sizes $N\equiv h^{(r)}+h^{(\ell)}\pmod2$ and has an effective conformal weight
	\begin{equation*}
		X_\text{eff}\left(h^{(\ell)},h^{(r)}\right) = -\frac16+\frac{\gamma}{\pi}\left(h^{(\ell)}+\frac12\right)^2 + \frac{\gamma}{\pi}\left( h^{(r)} + \frac12 -\frac{\pi}{2\gamma} \right)^2\,.
	\end{equation*}
	Numerical data from the exact diagonalization of small systems show that the multiplicities of these levels is $2(2h^{(\ell)}+1)(2h^{(r)}+1)$.  Given that Bethe states are $U_q(B_1)\otimes U_q(B_1)$ highest-weight states with left and right $U_q(B_1)$ spins $\mathcal{S}^{(\alpha)}=h^{(\alpha)}$, $\alpha=\ell,r$, this observation can be understood from self-duality and quantum group symmetries alone: each weight corresponds to a self-dual multiplet $[\mathcal{S}^{(\ell)},\mathcal{S}^{(r)}] \equiv (\mathcal{S}^{(\ell)}\otimes\mathcal{S}^{(r)}) \oplus (\mathcal{S}^{(r)} \otimes\mathcal{S}^{(\ell)})$ (hence $2(\mathcal{S} \otimes\mathcal{S})$ for $\mathcal{S}^{(\ell)}=\mathcal{S}^{(r)}=\mathcal{S}$) and can be written in the explicitly self-dual form
	\begin{align}
		\label{eq:ceff11-k}
		X_\text{eff}^{[A]}\left([\mathcal{S}^{(\ell)},\mathcal{S}^{(r)}]\right) &= -\frac16+\frac{\gamma}{4\pi}\left(2\mathcal{S}^<+1\right)^2 + \frac{\gamma}{4\pi}\left( 2\mathcal{S}^> + 1 -\frac{\pi}{\gamma} \right)^2\,,
	\end{align} 
	where $\mathcal{S}^<=\min(\mathcal{S}^{(\ell)},\mathcal{S}^{(r)})$, $\mathcal{S}^>=\max(\mathcal{S}^{(\ell)},\mathcal{S}^{(r)})$.  
	The conjectures (\ref{eq:ceff11-k}) together with extrapolations of our finite-size data \cite{D23.data} are presented in Fig.~\ref{fig:bc11_gs}.
	
	At the endpoints $\gamma=\pi/(2k+2)$, $k=2,3,4\dots$ of the intervals (\ref{eq:intervals11}) there is a level crossing between the ground states of the sectors $[\mathcal{S}^{(\ell)},\mathcal{S}^{(r)}]=[0,k]$ and $[0,k+1]$.  Hence, the effective central charge of the $D^{(2)}_3$ lattice model with boundary conditions $(\varepsilon,p)=(1,1)$ becomes
	\begin{equation}
		\label{eq:ceff11}
		c_\text{eff} = -24\,X_{\rm eff}^{(0)}=4-\frac{6\gamma}{\pi} - \frac{6\gamma}{\pi}\left(2\,\text{frac}\left(\frac{\pi}{2\gamma}\right) - 1 \right)^2\,.
	\end{equation}

	We expect that (\ref{eq:ceff11-k}) are the lower edges of continuous components of the spectrum of conformal weights: in the spin sectors shown in Figure~\ref{fig:bc11_gs} we have identified the root configurations of the first few (the lowest of expected towers) excitations that extrapolate to the  same  effective conformal weight in the scaling limit, see Table~\ref{tab:towerroots}.
	\begin{table}[t]
		\begin{center}
			\begin{tabular}
				{ccccc}
				\hline\\[-12pt]
				$[\mathcal{S}^{(\ell)},\mathcal{S}^{(r)}]$ & ${\tt n}$ & no.\ of & \multicolumn{2}{c}{additional roots on} \\
				&&(\ref{d23-strings})-strings & level $1$ & level $2$\\\hline\hline\\[-12pt]
				$[0,1]$ & $0^*$ & $(N-1)/2$  & -- & --\\
				& $1^*$ & $(N-2)/2$ & $x^{[1]}$  & $0$\\
				& $2$ & $(N-3)/2$ & $x_1^{[1]}$, $x^{[1]}_2+\ri\pi/2$ & $x^{[2]}$, $x^{[2]}+\ri\pi$\\
				& $3$ & $(N-4)/2$ & $x_1^{[1]}$, $x^{[1]}_k+\ri\pi/2$, $k=2,3$ & $0$, $x^{[2]}$, $x^{[2]}+\ri\pi$\\\hline\\[-12pt]
				$[0,2]$ & $0$ & $(N-2)/2$ & -- & -- \\
				& $1$ & $(N-3)/2$  & $x^{[1]}+\ri\pi/2$  & $0$\\
				& $2$ & $(N-4)/2$ & $x_k^{[1]}+\ri\pi/2$, $k=1,2$ & $x^{[2]}$, $x^{[2]}+\ri\pi$\\
				& $3$ & $(N-5)/2$ & $x_k^{[1]}+\ri\pi/2$, $k=1,2,3$ & 0, $x^{[2]}$, $x^{[2]}+\ri\pi$\\\hline\\[-12pt]
				$[0,3]$ & $0$ & $(N-3)/2$  & -- & --\\
				& $1$ & $(N-4)/2$  & $x^{[1]}+\ri\pi/2$ & 0 \\\hline\\[-12pt]
				$[0,4]$ & $0$ & $(N-4)/2$ & -- & --\\\hline\\[-12pt]
				$[1,1]$ & $0^*$ & $(N-4)/2$ & $\ri\gamma$, $\ri\gamma$, $\ri\,y^{[1]}$ & $\pm \ri\,y^{[2]}$\\
				& $1^*$ & $(N-3)/2$  & $\ri\gamma$, $\ri\gamma$ & $0$\\
				& $2$ & $(N-4)/2$ & $\ri\gamma$, $\ri\gamma$, $x^{[1]}+\ri\pi/2$ & $x^{[2]}$, $x^{[2]}+\ri\pi$\\\hline\\[-12pt]
				$[1,2]$ & $0$ & $(N-3)/2$  & $x^{[1]}$ & -- \\
				& $1$ & $(N-4)/2$ & $x_1^{[1]}$, $x^{[1]}_2+\ri\pi/2$ & 0\\\hline
			\end{tabular}    
		\end{center}
		\caption{\label{tab:towerroots}Bethe root configurations for the lowest class A states around $\gamma=\pi/5$: $x^{[a]}_k$ and $y^{[a]}_k$ take real values. In the scaling limit, the conformal weights with the same  $U_q(B_1)$ spins $[\mathcal{S}^{(\ell)},\mathcal{S}^{(r)}]$ (but different ${\tt n}$) become degenerate, indicating the emergence of a continuous spectrum.  An exception are the lowest spin $[\mathcal{S}^{(\ell)}=1,\mathcal{S}^{(r)}]$ states (with labels ${\tt n}=0^*$ and $1^*$): they are no longer part of the continuum but transmuted into pairs (realized for even and odd $N$, respectively) of discrete states for this value of $\gamma$.}
	\end{table}
	For the sector $[\mathcal{S}^{(\ell)},\mathcal{S}^{(r)}]=[0,2]$ we have visualized the mechanism for building this tower of excitations in Fig.~\ref{fig:bc11_towerroots}:
	\begin{figure}[ht]
		\centerline{\includegraphics[width=0.9\textwidth]{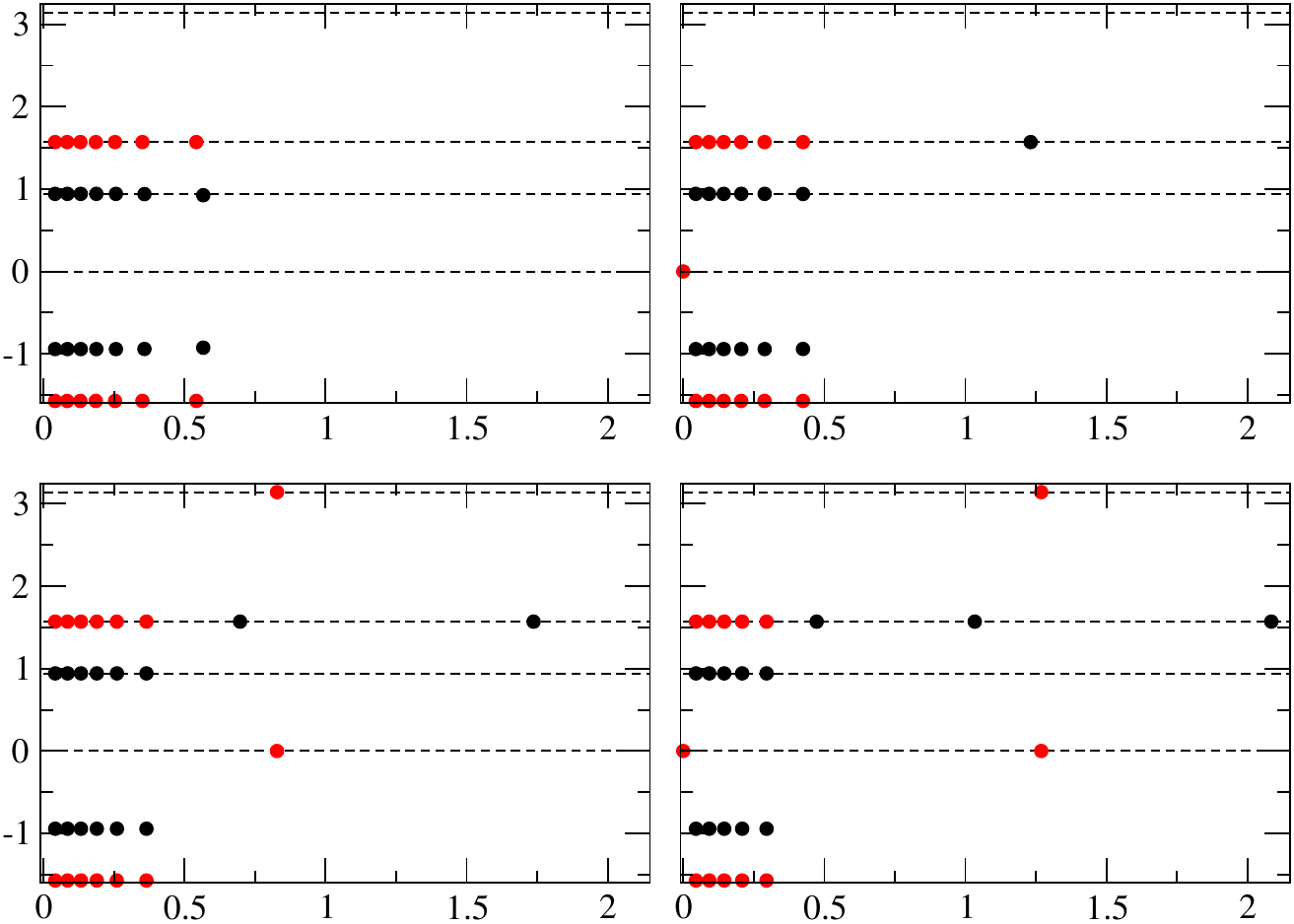}}
		\caption{Root configurations $\{\BR{a}{k}\}$ for the $[\mathcal{S}^{(\ell)},\mathcal{S}^{(r)}]=[0,2]$ levels shown in Fig.~\ref{fig:bc11_tower}: ground state for $N=16$ (top left), ground state for $N=15$ (top right), first excitation in the continuum for $N=16$ (lower left) and for $N=15$ (lower right). Black (red) dots are first (second) level roots, the dashed lines are at $\Im m(\BR{a}{k})=0$, $\frac{\pi}{2}-\gamma$, $\frac{\pi}{2}$ and $\pi$.
			\label{fig:bc11_towerroots}}
	\end{figure}
	in the center of the interval  $\pi/6<\gamma<\pi/4$ it is similar as in the periodic $D^{(2)}_3$ model. A pair of 2-strings (\ref{d23-strings}) is  replaced by level-1 roots with imaginary part $\pi/2$ and two level-2 roots with the same real part and imaginary parts $0$ and $\pi$, respectively (plus a single root $\BR{2}{0}=0$ in the excitations for chains of length $N\equiv \mathcal{S}^{(r)}+ \mathcal{S}^{(\ell)}+1\pmod2$). This procedure can be repeated $\mathcal{O}(N)$ times, i.e.\ until no strings (\ref{d23-strings}) are left, resulting in a tower of states extending beyond the low energy regime.\footnote{In the periodic model the ``new" second level roots can be freely distributed on the real line and with $\Im m(\BR{2}{j})=\pi$ which leads to the appearance of \emph{two} continua related by the $Z_2$ symmetry (\ref{Z2symmetry}).  Here a pair of roots $\{u^{[2]},u^{[2]}+\ri\pi\}$ is mapped onto itself.}
	Note that the root configurations change for values of $\gamma$  outside the interval where the lowest of these states is the ground state.

\begin{figure}[ht]
		\centerline{\includegraphics[width=0.8\textwidth]{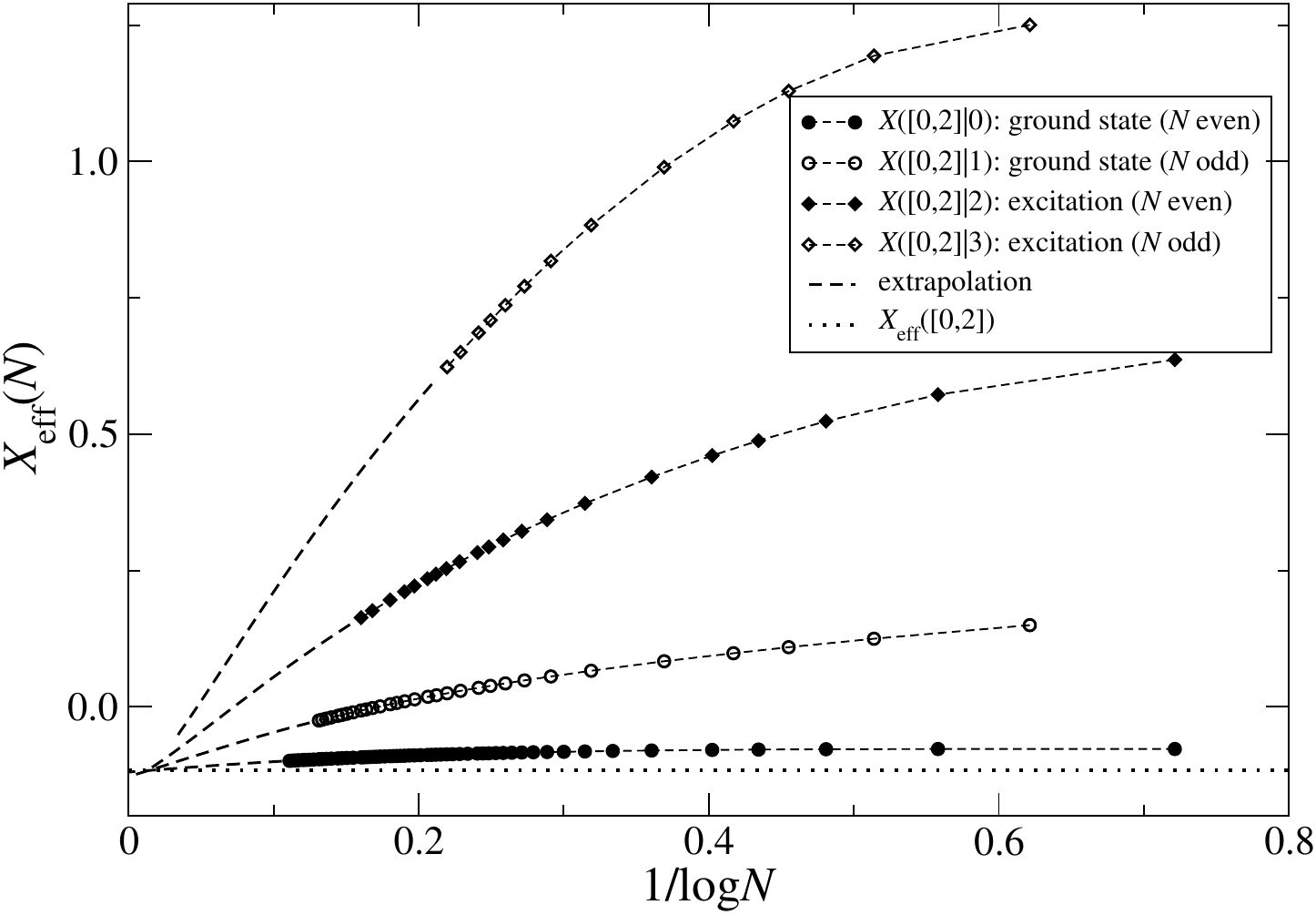}}
		\caption{Scaling of the effective conformal weights $X_{\text{eff}}(N)=\frac{N}{\pi v_F}\left(E(N)-N e_\infty-f_\infty\right)$ of the  lowest spin $[\mathcal{S}^{(\ell)},\mathcal{S}^{(r)}]=[0,2]$ levels for $\gamma=\frac15 \pi$: extrapolation gives the same value (\ref{eq:ceff11-k}) in the thermodynamic limit but with different subleading (logarithmic) corrections (\ref{eq:subleadingamps}). 
			\label{fig:bc11_tower}}
	\end{figure}

	As shown in Fig.~\ref{fig:bc11_tower}, there are strong corrections to scaling to the corresponding conformal weights.
	Analyzing these corrections in the $[\mathcal{S}^{(\ell)},\mathcal{S}^{(r)}]=[0,2]$-tower, we find that they have both logarithmic and power-law contributions.  Based on our numerical data \cite{D23.data}, we conjecture that the amplitudes of the logarithmic (leading) contributions are
	\begin{align}
		\label{eq:subleadingamps}
		X_{\text{eff}}\left([0,2]|{\tt n}\right) &= X_\text{eff}^{[A]}\left([0,2]\right) + {(2{\tt n}+1)^2}\,\frac{A(\gamma)}{\log^2(N/N_0)}
		+\mathcal{O}\left(N^{-x}\right)\,\quad\text{for~}{\tt n}=1,2,3,\dots\,,
	\end{align}
	while further contributions vanish as power laws with a $\gamma$-dependent exponent $x>0$.  Such terms are expected to be generated by the presence of perturbations to the fixed-point Hamiltonian by irrelevant operators in the lattice model \cite{Card86a,SiBo06,LSUO24}. In all states considered the exponent $x$ appears to vanish as $\gamma\to\pi/4$, indicating that the perturbation becomes marginally irrelevant in this limit.
	
	The amplitude of the logarithmic corrections to the effective weight $X_{\text{eff}}\left([0,2]|{\tt n}=0\right)$ of the lowest state in the continuum is too small to separate it from the power laws in the numerical data.  In the differences $\Delta_{\tt n}=X_{\text{eff}}\left([0,2]|{\tt n}+2\right)-X_{\text{eff}}\left([0,2]|{\tt n}\right)$, however, the amplitudes of the latter can be partially canceled. Based on their analysis we conjecture that (\ref{eq:subleadingamps}) holds for ${\tt n}=0$, too.
	
	Following the finite-size scaling behavior of the weights $X_\text{eff}([0,\mathcal{S}^{(r)}]|{\tt n})$ with ${\tt n}=0,1$ beyond the range (\ref{eq:intervals11}), i.e.\ to anisotropies $\gamma\lesssim\pi/(2\mathcal{S}^{(r)}+2)$, we observe a non-analytic change in the $\gamma$-dependence. At the same value of $\gamma$, the logarithmic corrections to scaling disappear and the subleading terms become pure power laws. A similar behaviour has been found in the $D^{(2)}_2$ (or staggered six-vertex) model, where this transmutation has been related to the appearance of the discrete levels in the spectrum of the 2D black hole CFT \cite{FrGe22}. 
	Following Eq.~(5.18) of Ref.~\cite{Frahm:2023lpe}, we conjecture the corresponding effective conformal weight to be
	\begin{subequations}
		\label{eq:xeff11_discrete}
		\begin{equation}
			\begin{aligned}
				X_{\text{eff}}^*{([0,\mathcal{S}^{(r)}])} &= X_{\text{eff}}^{[A]}([0,\mathcal{S}^{(r)}])
				- \frac{\gamma}{\pi-2\gamma}\left(\mathcal{S}^{(r)}+1-\frac\pi{2\gamma}\right)^2\,
				\qquad\text{for~}\gamma<\frac{\pi}{2\mathcal{S}^{(r)}+2}\,.
			\end{aligned}
		\end{equation}
		Upon lowering $\gamma$ further, additional discrete levels emerge from the continuum, e.g.\
		\begin{equation}\label{gnjdjdne}
			\begin{aligned}
				X_{\text{eff}}^{**}{([0,\mathcal{S}^{(r)}])} &= X_{\text{eff}}^{[A]}([0,\mathcal{S}^{(r)}])
				- \frac{\gamma}{\pi-2\gamma}\left(\mathcal{S}^{(r)}+2-\frac\pi{2\gamma}\right)^2\,
				\qquad\text{for~}\gamma<\frac{\pi}{2\mathcal{S}^{(r)}+4}\,.
			\end{aligned}
		\end{equation}
	\end{subequations}
	These levels are also displayed in Fig.~\ref{fig:bc11_gs}.
	Similar transmutations of the lowest continuum levels into discrete ones appear in the highest-weight class~A states of all Cartan sectors $0\leq h^{(\ell)} \leq h^{(r)}$.
	
	Summarizing our results for the class~A states: we have identified towers of conformal weights separated by gaps $\propto 1/\log^2(N/N_0)$ such as (\ref{eq:subleadingamps}) which realize continuous components of the conformal spectrum of the underlying theory starting at values (\ref{eq:ceff11-k}) in the scaling limit.
	\begin{subequations}
		\label{eq:xeff11-slsr}
		Rewriting these lower edges as
		\begin{equation}
			\label{eq:x11-Aconti}
			\begin{aligned}
				X_\text{eff}^{[A]}\left([\mathcal{S}^{(\ell)},\mathcal{S}^{(r)}]\right) =&
				-\frac1{12} + \frac{\gamma}{4\pi}\left( 2\mathcal{S}^> + 1 -\frac{\pi}{\gamma} \right)^2\\
				&
				-\frac1{12}+\frac{\gamma}{4\pi}\left(2\mathcal{S}^<+1\right)^2\,,
			\end{aligned}
		\end{equation}
		the first line coincides with the lower edge of the continuous parts of the spectrum of the non-compact black hole boundary CFT as realized in the $D^{(2)}_2$ model with $U_q(B_1)$ quantum group symmetry for $\varepsilon=1$ (or one half of the CFT describing the $D^{(2)}_3$ model with periodic boundary conditions \cite{Frahm:2023lpe}), see Eq.~(\ref{eq:D22cont}).  
		The second line in (\ref{eq:x11-Aconti}) is the contribution of another (compact) critical degree of freedom with conformal weights as in (\ref{eq:Xeff01}) for boundary conditions $(\varepsilon,p)=(0,1)$ in the universality class of the afm Potts model with free boundaries.
		
		Similarly, we propose that (\ref{eq:xeff11_discrete}) are the first two in a sequence of discrete levels emerging from the continuum:\footnote{Note that $X_\text{eff}^{[A]}\left([\mathcal{S}^{(\ell)},\mathcal{S}^{(r)}]|a\right)\to-\frac16+a$ for $\gamma\to0$ independent of $\mathcal{S}^{(\ell,r)}$.}
		\begin{equation}
			\label{eq:x11-Adiscrete}
			\begin{aligned}
				X_\text{eff}^{[A]}\left([\mathcal{S}^{(\ell)},\mathcal{S}^{(r)}]|a\right) =&-\frac1{12} + \frac{\gamma}{4\pi}\left(2\mathcal{S}^>+1-\frac{\pi}{\gamma}\right)^2 
				+\frac{\gamma}{\pi-2\gamma}\,s_a^2\\
				& -\frac1{12} + \frac{\gamma}{4\pi}\left(2\mathcal{S}^<+1\right)^2 \,,\\
				\text{with~}&s_a=\pm\ri \left(\mathcal{S}^>+1+a-\frac{\pi}{2\gamma}\right)\,,\quad a=0,1,2,\dots<\frac\pi{2\gamma}-(\mathcal{S}^>+1)\,.
			\end{aligned}
		\end{equation}
	\end{subequations}
	Here the contribution of the compact critical degree of freedom (second line) is unchanged while the first line of (\ref{eq:x11-Adiscrete}), together with the restriction on the possible values of $a$, can be identified with the spectrum of discrete states of the black hole boundary CFT allowed by the unitarity condition.  This provides further support for our interpretation of the finite-size spectrum from the class~A states.
	
	One might speculate that the factor multiplying $s_a^2$ also enters in the amplitudes $A(\gamma)$ of the log.\ corrections of the continuum states (\ref{eq:subleadingamps}).  This is not easy to verify, however, due to the (possibly strong) power law contributions coming from the second degree of freedom.\newline

	In addition to the class~A levels discussed above, we have identified the conformal weights of several other discrete states with purely power law corrections to scaling.  These states are characterized by Bethe root configurations containing one or two singular roots (\ref{sing}) on level one, i.e.\ $n_{\text{sing}}^{[1]}\neq0$, and the corresponding weights do not appear in the discrete part of the spectrum of the 2D black hole CFT.
	Moreover, our numerical data for small systems indicate that the multiplicities of these levels is in general larger that those for the class A states: for $\mathcal{S}^{(r)}\neq \mathcal{S}^{(\ell)}$ their representation content is $k\,[\mathcal{S}^{(\ell)},\mathcal{S}^{(r)}] = k\left((\mathcal{S}^{(\ell)}\otimes\mathcal{S}^{(r)}) \oplus (\mathcal{S}^{(r)} \otimes\mathcal{S}^{(\ell)})\right)$ with an even integer $k$.
	
	\paragraph{Class B: additional discrete states with one singular root on level 1 ($n_{\text{sing}}^{[1]}=1$).}
	
\begin{figure}[t]
		\centerline{\includegraphics[width=0.8\textwidth]{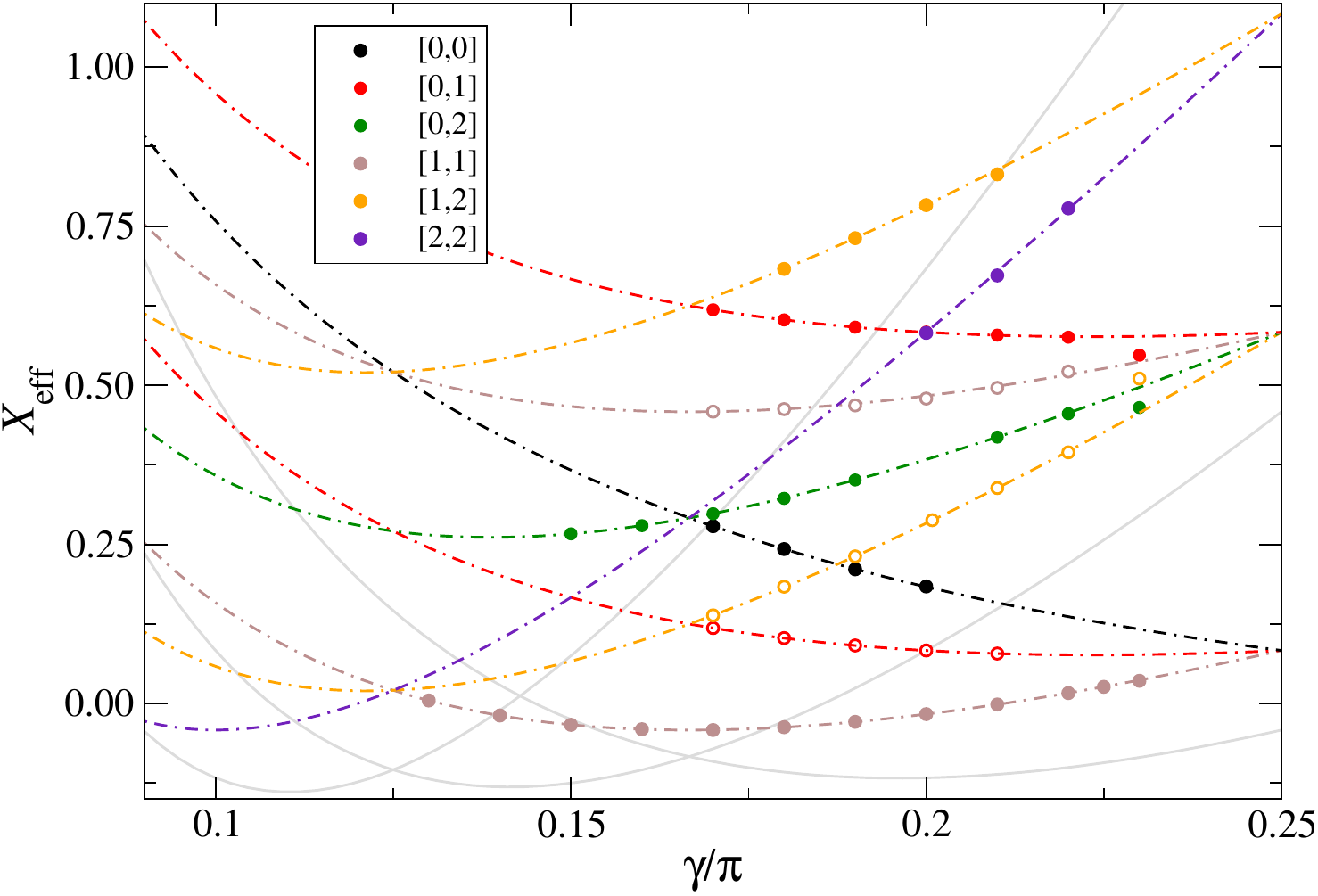}}
		\caption{Conjectured effective conformal weights $X_{\text{eff}}^{[B]}$ for the class~B states from the discrete spectrum, see (\ref{eq:xeff11_other}). Colors indicate the sector $[\mathcal{S}^{(\ell)},\mathcal{S}^{(r)}]$ as shown in the legend box, filled (open) circles are extrapolations of the numerical data \cite{D23.data} for even (odd) length lattices. Grey lines indicate the lower edges of the $[0,\mathcal{S}^{(r)}]$-continua.
			\label{fig:bc11_other}}
	\end{figure}

	The lowest-energy states with one singular first-level root (\ref{sing}) and $U_q(B_1)$ spins $\mathcal{S}^{(\ell,r)}$ have conformal weights
	\begin{equation}
		\label{eq:xeff11_other}
		\begin{aligned} 
			X_\text{eff}^{[B]}\left([\mathcal{S}^{(\ell)},\mathcal{S}^{(r)}]\right) = X_\text{eff}^{[B]}\left([\mathcal{S}^{(r)},\mathcal{S}^{(\ell)}]\right) =
			-\frac16 + \frac{\gamma}{4\pi}\,\sum_{\alpha=\ell,r}\left(2\mathcal{S}^{(\alpha)}+1-\frac{\pi}{2\gamma}\right)^2+\frac18\,,
		\end{aligned}
	\end{equation}
	see Fig.~\ref{fig:bc11_other} for the states with $\mathcal{S}^{(\alpha)}\leq 2$.
	Also shown are some excited class B states with conformal weights
	\begin{equation*}
		X_\text{eff}^{[B]}\left([\mathcal{S}^{(\ell)},\mathcal{S}^{(r)}]\right)+\frac12\,.
	\end{equation*}
	For $\gamma\to\pi/4$, i.e.\ at the boundary of the critical regime considered in this paper, the conformal weights $X^{[B]}_{\text{eff}}$ approach $-\frac16+\frac{2k+1}4$ with non-negative integers $k$. Such values are absent in the conformal spectrum obtained from the class~A states.  Therefore, they cannot be interpreted in the context of the black hole CFT.

	\paragraph{Class C: additional discrete states with two singular roots on level 1 ($n_{\text{sing}}^{[1]}=2$).}
	
	Finally, we have identified a class of primaries and descendants with conformal weights depending linearly on $\gamma$,
	\begin{equation}
		\label{eq:xeff11-linear}X_\text{eff}^{[C]}\left([\mathcal{S}^{(\ell)},\mathcal{S}^{(r)}]\right) = -\frac16 + \frac{\gamma}{4\pi}\,\sum_{\alpha=\ell,r}\left(2\mathcal{S}^{(\alpha)}+1\right)^2+{\tt d}\,.
	\end{equation}
	This coincides with the operator content (\ref{eq:Xeff01}) of the model with boundary conditions $(\varepsilon,p)=(0,1)$ which we have identified with that of two afm Potts models with free boundaries.
	All of these states have $n_{sing}^{[1]}=2$ singular first level roots (\ref{sing}). The corrections to scaling are pure power laws indicating that they do not belong to a continuous part of the conformal spectrum.

	\section{Discussion}\label{sec:discussion}
	
	We have determined the spectrum of the transfer matrices and corresponding Bethe equations for the quantum-group-invariant $D^{(2)}_{n+1}$ models with $\varepsilon=1$ using the analytical Bethe ansatz. Focusing on the case $n=2$, we have explicitly constructed the spin-chain Hamiltonians for the various boundary conditions $(\varepsilon,p)$, which is nontrivial for the case $(1,1)$ as it involves the second derivative of the transfer matrix.
	
	We find that the Hamiltonian is critical in the parametric regime \eqref{mfsoleme} for all four boundary conditions $(\varepsilon,p)$, just as for PBCs.  We utilize our Bethe ansatz to investigate the finite-size spectrum of the $D^{(2)}_3$ model for large system sizes $N\gg 1$.
	In our previous study of the periodic model this approach led to the identification of two critical non-compact degrees of freedom, each described by the 2D black hole CFT in the scaling limit. Our numerical data indicate that imposing different quantum group invariant boundary conditions on the lattice model has a profound effect on its properties: for $\varepsilon=1$ the continuous symmetries of the model, $U_q(B_2)$ for $p=0$ and $U_q(B_1)\otimes U_q(B_1)$ for $p=1$, are spontaneously broken.  Moreover, we find signatures of a single non-compact degree of freedom only for BCs with $\varepsilon=1$, while the spectrum of conformal weights is purely discrete when $\varepsilon=0$. The disappearance of the continuous part of the spectrum has also been observed in the staggered six-vertex model. Assuming that the scaling limit of the periodic $D^{(2)}_{n+1}$ chains is described by non-compact CFTs for all $n$ (as appears to be the case for the $A^{(2)}_{n+1}$ series of lattice models \cite{VeJS14, VeJS16a}) this leads us to conjecture that under quantum group invariant BCs (\ref{KRb}), (\ref{KL}) only those with $\varepsilon=1$ lead to a continuous spectrum of conformal weights corresponding to non-compact branes.  This could be checked based on our construction of the Bethe equations for the rank $n$ chains in Section~\ref{subsec:BA}.
	
	The conformal spectrum of the self-dual $U_1(B_1)\otimes U_q(B_1)$-symmetric $D^{(2)}_3$ models, i.e.\ with $p=1$, allows for a partial interpretation when compared to the $U_q(B_1)$-invariant $D^{(2)}_2$ chain: in the scaling limit the latter is known to be in the universality class of the (compact) afm Potts model with free boundaries for $\varepsilon=0$, while its low-energy effective description for $\varepsilon=1$ is a 2D black hole boundary CFT.  Here we have found for BCs $(\varepsilon,p)=(0,1)$ that the conformal weights (\ref{eq:Xeff01}) contain contributions of two
	afm Potts models.  For $(\varepsilon,p)=(1,1)$ the situation is more complicated, but part of the conformal spectrum can be decomposed into contributions of an afm Potts model and a black hole boundary CFT, see (\ref{eq:xeff11-slsr}).  This supplements our earlier observation for the periodic model \cite{Frahm:2023lpe}.
	
	Further insights into the scaling limit of the $D^{(2)}_3$ model and the possible conformal boundary conditions might be gained by considering the spin chains with different choices of the parameters $(\varepsilon,p)$ in the K-matrices (\ref{KRb}) and (\ref{KL}) on the two ends.  Moreover, we have restricted our attention in this paper on values of the anisotropy parameter $\gamma$ in the domain (\ref{mfsoleme}).  We expect that the system remains critical for any real values of $\gamma$, so it might be worthwhile to explore the finite-size spectrum in other domains.
	
	\section*{Acknowledgements}
	
	HF and SG acknowledge partial funding provided by the Deutsche Forschungsgemeinschaft (DFG) under grant No. Fr 737/9-2 as part of the research unit \emph{Correlations in Integrable Quantum Many-Body Systems} (FOR2316). 
	SG received further funding by the EPSRC under grant EP/X030881/1.
	RN was supported in part by the National Science Foundation under grant PHY 2310594, and by a Cooper fellowship.
	ALR was supported by Grant 404 No. 18/EPSRC/3590 and by the UKRI Future Leaders Fellowship (grant number MR/T018909/1). 
	We thank the following workshops and/or institutions, where part of this work was performed: the mathematical research institute MATRIX in Australia (RN and ALR), the University of Miami (ALR),  Durham University (SG) and the Student Workshop on Integrability at E\"otv\"os Lor\'and University (SG and RN).
	HF and SG thank Gleb A. Kotousov for his helpful assistance with the root-density approach in the open case; and ALR and SG thank Simon Ekhammar for interesting discussions.

	\appendix

\section{$D_{n+1}^{(2)}$ R-matrix}\label{sec:Rmat}

We use the $D_{n+1}^{(2)}$ R-matrix given by Jimbo 
\cite{Jimbo:1985ua}, except we use the
variables $u$ and $\eta$ instead of $x$ and $k$, respectively, which are related as follows:
\begin{equation}
x = e^u \,, \qquad \qquad k = e^{2 \eta} \,.
\end{equation}
We also multiply the Jimbo R-matrix by an overall factor $e^{-2u}\, 
e^{-2(n+1)\eta}$ in order to have nice crossing and 
unitarity properties. (See also \cite{Bazhanov:1984gu, 
Bazhanov:1986mu}.) Hence, this R-matrix is given,
as in Appendix A of \cite{Nepomechie:2017hgw}, by
\begin{equation}
R(u)=e^{-2u}\,e^{-2(n+1)\eta}R_J(u)
\label{RD2}
\end{equation}
with
\begin{align}
R_J(u)= & \left(e^{2u}-e^{4\eta}\right)\left(e^{2u}-e^{4n\eta}\right)\sum_{\alpha\neq n+1,n+2}e_{\alpha\alpha}\otimes e_{\alpha\alpha}
+e^{2\eta}\left(e^{2u}-1\right)\left(e^{2u}-e^{4n\eta}\right)
\sum_{\substack{\alpha\neq\beta,\beta'\\\alpha\,\textrm{or}\,\beta\neq n+1,n+2}}
\nonumber\\
& \cdot e_{\alpha\alpha}\otimes e_{\beta\beta}-
\left(e^{4\eta}-1\right)
\left(e^{2u}-e^{4n\eta}\right)
\left(\sum_{\substack{\alpha<\beta,\alpha\neq\beta' \\
\alpha,\beta\neq n+1,n+2}}+e^{2u}\sum_{\substack{\alpha>\beta,\alpha\neq \beta'\\
\alpha,\beta\neq n+1,n+2}}\right)
e_{\alpha\beta}\otimes e_{\beta\alpha}
\nonumber\\
& -\frac{1}{2}
\left(e^{4\eta}-1\right)\left(e^{2u}-e^{4n\eta}\right)
\Bigg(
\left(e^{u}+1\right)
\left(
\sum_{\alpha<n+1,\beta=n+1,n+2}+e^{u}\sum_{\alpha>n+2,\beta=n+1,n+2}
\right)
\non\\
&
\cdot\left(
e_{\alpha\beta}\otimes e_{\beta\alpha}+e_{\beta'\alpha'}\otimes e_{\alpha'\beta'}
\right)
+
\left(e^{u}-1\right)
\left(
-\sum_{\alpha<n+1,\beta=n+1,n+2}+e^{u}\sum_{\alpha>n+2,\beta=n+1,n+2}
\right)
\non\\
&
\cdot
\left(
e_{\alpha\beta}\otimes e_{\beta'\alpha}+e_{\beta'\alpha'}\otimes e_{\alpha'\beta}
\right)\Bigg)+
\sum_{\alpha,\beta\neq n+1,n+2}a_{\alpha\beta}(u)e_{\alpha\beta}\otimes e_{\alpha'\beta'}+
\frac{1}{2}\sum_{\alpha\neq n+1,n+2,\beta=n+1,n+2}
\non\\
&
\cdot
\left(
b_\alpha^{+}(u)
\left(
e_{\alpha\beta}\otimes e_{\alpha'\beta'}+e_{\beta'\alpha'}\otimes e_{\beta\alpha}
\right)
+
b_\alpha^{-}(u)
\left(
e_{\alpha\beta}\otimes e_{\alpha'\beta}+e_{\beta\alpha'}\otimes e_{\beta\alpha}
\right)
\right)
\non\\
&
+\sum_{\alpha=n+1,n+2}
\left(c^{+}(u)e_{\alpha\alpha}\otimes e_{\alpha'\alpha'}+
c^{-}(u)e_{\alpha\alpha}\otimes e_{\alpha\alpha}\right.
\non\\
&
\left. +\,
d^{+}(u)e_{\alpha\alpha'}\otimes e_{\alpha'\alpha}+
d^{-}(u)e_{\alpha\alpha'}\otimes e_{\alpha\alpha'}
\right)\,,
\label{RmatD2}
\end{align}
where for $\alpha,\beta\neq n+1,n+2$
\begin{equation}
a_{\alpha\beta}(u)=\begin{cases}
(e^{4\eta}e^{2u}-e^{4n\eta})(e^{2u}-1)& \alpha=\beta\\
(e^{4\eta}-1)(e^{4n\eta}e^{2\eta(\bar\alpha-\bar\beta)}(e^{2u}-1)-\delta_{\alpha\beta'}(e^{2u}-e^{4n\eta}))& \alpha<\beta\\
(e^{4\eta}-1)e^{2u}(e^{2\eta(\bar\alpha-\bar\beta)}(e^{2u}-1)-\delta_{\alpha\beta'}(e^{2u}-e^{4n\eta}))& \alpha>\beta
\end{cases}\,,
\end{equation}
\begin{equation}
b_{\alpha}^{\pm}(u)=\begin{cases}
\pm e^{2\eta(\alpha-1/2)}(e^{4\eta}-1)(e^{2u}-1)(e^u\pm e^{2n\eta})& \alpha<n+1\\
e^{2\eta(\alpha-n-5/2)}(e^{4\eta}-1)(e^{2u}-1)e^u(e^u\pm e^{2n\eta})& \alpha>n+2
\end{cases}\,,
\end{equation}
\begin{equation}
c^{\pm}(u)=\pm\frac{1}{2}(e^{4\eta}-1)(e^{2n\eta}+1)e^u(e^u\mp 1)(e^u\pm e^{2n\eta})+e^{2\eta}(e^{2u}-1)(e^{2u}-e^{4n\eta})\,,
\end{equation}
\begin{equation}
d^{\pm}(u)=\pm\frac{1}{2}(e^{4\eta}-1)(e^{2n\eta}-1)e^u(e^u\pm 1)(e^u\pm e^{2n\eta})\,,
\end{equation}
and
\begin{equation}
\bar{\alpha}=\begin{cases}
\alpha+1 & 1\le\alpha<n+1\\
n+\frac{3}{2} & \alpha=n+1\\
n+\frac{3}{2} & \alpha=n+2\\
\alpha-1 & n+2<\alpha\le 2n+2
\end{cases} \,,
\label{alphabarD2}
\end{equation}
\be
\alpha'=2n+3-\alpha\,.
\ee
The elementary matrices $e_{\alpha\beta}$ have dimension $(2n+2) \times (2n+2)$ with
\begin{equation}
\alpha,\beta=1,\dots,2n+2\,.
\end{equation}

This R-matrix has the crossing symmetry
\begin{equation}
R_{12}(u)=V_1\, R_{12}^{t_2}(-u-\rho)\, V_1
= V_2^{t_2}\, R_{12}^{t_1}(-u-\rho)\, V_2^{t_2} \,,
\label{crossing}
\end{equation}
with the crossing parameter $\rho = -2 n\eta$. 
The crossing matrix $V$ is given by
\begin{equation}
V= \sum_{k=1}^{2n+2} v_{k}\, e_{k,2n+3-k} \,, 
\end{equation}
where
\begin{align} 
v_{k} &= \left\{ \begin{array}{cl}
e^{-(2n+1-2k)\eta} & k = 1, \ldots, n \\
1 & k = n+1, n+2 \\
e^{-(2n+5-2k)\eta} & k = n+3, \ldots , 2n+2
\end{array} \right.  \,,
\end{align}
and satisfies $V^2 = \id$. The corresponding matrix $M$ is defined by
\begin{equation}
M =  V^{t}\, V\,, 
\label{Mdef}
\end{equation}
and is given by a diagonal matrix
\begin{equation}
M  =
\diag(e^{4(n+\frac{3}{2}-\bar\alpha)\eta})\,, 
\qquad \alpha = 1, 2, \ldots \,, 2n+2 \,,
\label{Mmat}
\end{equation}
where $\bar\alpha$ is defined by (\ref{alphabarD2}).
	
	\section{Quantum group and duality symmetries}\label{ap:QGandDuality}
	
	We provide here further details about the quantum group and duality symmetries of the transfer matrix. The explicit form of the quantum group generators and their coproducts are given in Section \ref{subap:generators}. The definition of the duality operator and its action on the quantum group generators are presented in Section \ref{subap:dualityandgenerators}.
	
	\subsection{Generators and their coproducts}\label{subap:generators}
	
	It was shown in \cite{Nepomechie:2018wzp} that the transfer matrix 
	$t(u, \varepsilon ,p)$ \eqref{transfer}
	has the $U_q\left(B_{n-p}\right)\otimes U_q\left(B_p\right)$ quantum group symmetry, see \eqref{QGsym}. In order to define the quantum group generators,
	we recall that the $D_{n+1}^{(2)}$ generators
	in the fundamental representation corresponding to simple roots can be written as
	\begin{align}
		& H_j=e_{j,j}-e_{2n+3-j,2n+3-j}, && j=1,...,n, \nonumber\\
		& E_0^+=\frac{(-1)^n}{\sqrt{2}}\left(e_{n+1,1}-e_{n+2,1}+e_{2n+2,n+1}-e_{2n+2,n+2}\right),\nonumber\\
		& E_{j}^{+}=e_{j,j+1}+e_{2n+2-j,2n+3-j}, && j=1,...,n-1,\nonumber\\
		& E_{n}^{+}=\frac{1}{\sqrt{2}}\left(e_{n,n+1}+e_{n,n+2}-e_{n+2}-e_{n+2,n+3}-e_{n+1,n+3}\right),\nonumber\\
		& E_j^{-}=\left(E_j^+\right)^t,\label{eq:unbroken generators}
	\end{align}
	where $(e_{i,j})_{a,b}=\delta_{i,a}\delta_{j,b}$. 
	
	A key point is that the $p$-th generator $E_{p}^{\pm}$
	is broken; hence, the remaining symmetry is described by the $n$ remaining generators that form  the ``left" and ``right" algebras.
	There are $n-p$ generators of the ``left" algebra, denoted by a superscript $(\ell)$, which are given by
	\begin{equation}
		H_{j}^{(\ell)}(p)=H_{p+j}, \quad E_{j}^{\pm\,(\ell)}(p)=E_{p+j}^{\pm}, \quad j=1,..., n-p\,;
		\label{eq:leftgenerators}
	\end{equation}
	and there are $p$ generators of the ``right" algebra, denoted by a superscript $(r)$, which can be written as
	\begin{equation}
		H_{j}^{(r)}(p)=H_{p-j+1}, \quad E_{j}^{\pm\,(r)}(p)=E_{p-j}^{\mp}, \quad j=1,...,p \,.
		\label{eq:rightgenerators}
	\end{equation}
	We note that the definition of the ``right" generators
	in \eqref{eq:rightgenerators} differs from the one
	in Eqs. (B.7) in \cite{Nepomechie:2018dsn} and
	(A.6) in \cite{Nepomechie:2018wzp}). While the two definitions are equivalent, the new choice ensures non-negative values for the highest-weight-state eigenvalues $h_{j}^{(r)}$, see \eqref{CartanCardinalities2}.
	
	The ``left" and ``right" algebras are given by the following commutation relations
	
	\begin{align}
		&  \left[H_i^{(\ell)}(p),H_j^{(\ell)}(p)\right]=0 \,,  && \left[H_i^{(r)}(p),H_j^{(r)}(p)\right]=0 \,,\nonumber\\
		&  \left[H_i^{(\ell)}(p),E^{\pm\, (\ell)}_j(p)\right]=\pm \alpha_i^{(j)}E^{\pm\, (\ell)}_j(p) \,,  && \left[H_i^{(r)}(p),E^{\pm\, (r)}_j(p)\right]=\pm \alpha_i^{(j)}E^{\pm\, (r)}_j(p) \,,\\
		& \left[E_i^{+\,(\ell)}(p),E_j^{-\,(\ell)}(p)\right]=\delta_{i,j}\sum_{k=1}^{n-p}\alpha_k^{(j)}H_k^{(\ell)}(p) \,, && \left[E_i^{+\,(r)}(p),E_j^{-\,(r)}(p)\right]=\delta_{i,j}\sum_{k=1}^{p}\alpha_k^{(j)}H_k^{(r)}(p) \,, \nonumber
	\end{align}
	where $\{\alpha^{(1)},...,\alpha^{(m)} \}$ are the simple roots of $B_{n-p}$ for $m=n-p$, and of $B_p$ for $m=p$. Their explicit form in the orthogonal basis is 
	\begin{align}
		& \alpha^{(j)}=e_j-e_{j+1}, \quad j=1,...,m-1 \,,\nonumber\\
		& \alpha^{(m)}=e_m \,,
	\end{align}
	where $(e_j)_a=\delta_{j,a}$.
	
	The 2-fold coproducts for the ``left"  generators are given, as in \cite{Nepomechie:2018wzp}, by
	\begin{align}
		& \Delta\left(H_j^{(\ell)}\right)=H_j^{(\ell)}\otimes \mathbb{I}+\mathbb{I}\otimes H_j^{(\ell)}\,, && j=1,...,n-p \,,\nonumber\\
		& \Delta\left(E_j^{\pm\,(\ell)}\right)=E_j^{\pm\,(\ell)}\otimes e^{(\eta+\ri \pi)H_j^{(\ell)}-\eta H_{j+1}^{(\ell)}}+ e^{-(\eta+\ri \pi)H_j^{(\ell)}+\eta H_{j+1}^{(\ell)}}\otimes E_j^{\pm \,(\ell)}\,, && j=1,...,n-p-1 \,,\nonumber\\
		& \Delta\left(E_{n-p}^{\pm\,(\ell)}\right)=E_{n-p}^{\pm\,(\ell)}\otimes e^{\eta H_{n-p}^{(\ell)}}+e^{-\eta H_{n-p}^{(\ell)}}\otimes E_{n-p}^{\pm\,(\ell)} \,,
	\end{align}
	and satisfy
	\begin{align}
		& \left[\Delta\left(H_i^{(\ell)}\right),\Delta\left(E_j^{\pm\,(\ell)}\right)\right]=\pm\alpha^{(j)}\Delta\left(E_j^{\pm\,(\ell)}\right) \,,\nonumber\\
		& \left[\Delta\left(E_i^{+\,(\ell)}\right),\Delta\left(E_j^{-\,(\ell)}\right)\right]_{\Omega^{(\ell)}}=\delta_{i,j}\frac{\sinh\left(2\eta\sum_{k=1}^{n-p}\alpha_k^{(j)}\Delta\left(H_k^{(\ell)}\right)\right)}{\sinh 2\eta} \,,
	\end{align}
	where
	\begin{equation}
		\Omega_{ij}^{(\ell)}=\begin{cases}
			e^{\ri\pi H_{\text{max}(i,j)}^{(\ell)}}\otimes \mathbb{I} \,, & |i-j|=1\, \text{and} \, 1 \le \text{min}(i,j)\le n-p-2 \\
			\mathbb{I}\otimes \mathbb{I} \,, & \text{otherwise}
		\end{cases} \,,
	\end{equation}
	and
	\begin{equation}  \left[A_i,B_j\right]_{\Omega^{(k)}}=\Omega_{ij}^{(k)}A_iB_j-B_jA_i\Omega_{ij}^{(k)} \,.
	\end{equation}
	
	The 2-fold coproducts for the ``right" generators are given by
	\begin{align}
		& \Delta\left(H_j^{(r)}\right)=H_j^{(r)}\otimes \mathbb{I}+\mathbb{I}\otimes H_j^{(r)} \,, && j=1,...,p \,, \nonumber\\
		& \Delta\left(E_j^{\pm\,(r)}\right)=E_j^{\pm\,(r)}\otimes e^{-(\eta+\ri \pi)H_j^{(r)}+\eta H_{j+1}^{(r)}}+ e^{(\eta+\ri \pi)H_j^{(r)}-\eta H_{j+1}^{(r)}}\otimes E_j^{\pm \,(r)} \,, && j=1,...,p-1 \,, \nonumber\\
		& \Delta\left(E_{p}^{\pm\,(r)}\right)=E_{p}^{\pm\,(r)}\otimes e^{-\eta H_{p}^{(r)}}+e^{\eta H_{p}^{(r)}}\otimes E_{p}^{\pm\,(r)} \,,
	\end{align}
	and satisfy
	\begin{align}
		& \left[\Delta\left(H_i^{(r)}\right),\Delta\left(E_j^{\pm\,(r)}\right)\right]=\pm\alpha^{(j)}\Delta\left(E_j^{\pm\,(r)}\right) \,, \nonumber\\
		& \left[\Delta\left(E_i^{-\,(r)}\right),\Delta\left(E_j^{+\,(r)}\right)\right]_{\Omega^{(r)}}=\delta_{i,j}\frac{\sinh\left(-2\eta\sum_{k=1}^{p}\alpha_k^{(j)}\Delta\left(H_k^{(r)}\right)\right)}{\sinh 2\eta} \,,
	\end{align}
	where
	\begin{equation}
		\Omega_{ij}^{(r)}=\begin{cases}
			e^{\ri\pi H_{\text{max}(i,j)}^{(r)}}\otimes \mathbb{I} \,, & |i-j|=1\, \text{and} \, 1 \le \text{min}(i,j)\le p-2\\
			\mathbb{I}\otimes \mathbb{I} \,, & \text{otherwise}
		\end{cases} \,,
	\end{equation}
	All ``left" generators commute with  ``right" generators.
	
	In order to construct $N$-fold coproducts, one simply uses coassociativity
	\begin{equation}
		(\mathbb{I} \otimes \Delta)\Delta=(\Delta\otimes \mathbb{I})\Delta 
	\end{equation}
	on the $(N-1)$-fold coproduct.
	
	\subsection{Duality}\label{subap:dualityandgenerators}
	
	It was shown in \cite{Nepomechie:2018wzp} that the transfer matrix $t(u, \varepsilon ,p)$ \eqref{transfer}
	has the $p \leftrightarrow n-p$ symmetry \eqref{duality}, where $\mathcal{U}$ is given by 
	\begin{equation}
		\mathcal{U} = U_1 \ldots U_N \,,  
	\end{equation}
	and $U$ is defined by
	\begin{equation}
		U=\begin{cases}
			\sum_{j=1}^{n}e_{j,n+2+j}+e_{n+1,n+1}-e_{n+2,n+2}-\sum_{j=1}^{n}e_{n+2+j,j}, & \text{for } \quad n \text{ even}\\
			\sum_{j=1}^{n}e_{j,n+2+j}+e_{n+1,n+2}-e_{n+2,n+1}-\sum_{j=1}^{n}e_{n+2+j,j}, & \text{for } \quad n \text{ odd}
		\end{cases} \,,
	\end{equation}
	and satisfies $U\, U^t = \id$.
	
	Duality maps ``right" generators to ``left" generators 
	\begin{align}
		&U\, H_i^{(r)}(p)\, U^{-1}=-H_i^{(\ell)}(n-p) \,,  \nonumber\\
		&U\, E_i^{\pm\,(r)}(p)\, U^{-1}= E_i^{\mp\,(\ell)}(n-p)\,, \quad \qquad i=1,...,p \,,
	\end{align}
	and vice-versa 
	\begin{align}
		&U\, H_i^{(\ell)}(p)\, U^{-1}=-H_i^{(r)}(n-p) \,, \nonumber\\
		&U\, E_i^{\pm\,(\ell)}(p)\, U^{-1}=\nu_i(p)\, E_i^{\mp\,(r)}(n-p) \,, \quad \qquad i=1,...,n-p \,,
	\end{align}
	where
	\begin{equation}
		\nu_i(p)=\begin{cases}
			-1 & n \text{ even and } i=n-p\\
			+1 & \text{otherwise}
		\end{cases} \,.
	\end{equation}
	The coproducts transform in a similar way.

	\section{The root-density approach in the open case:  deviations are important  \label{fmkdmngmocmf}}
	In this appendix, we demonstrate a small subtlety for the root-density approach when considering open BCs for string solutions whose imaginary part exactly cancels the additive shifts, in our case $\pm \ri \gamma$ ,$\pm 2\ri \gamma$, in the scattering phase. This requires more careful handling by taking into account the limit of vanishing deviations.  
	
	Let us consider as an example the case $(\varepsilon,p)=(0,0)$. The other BCs follow analogously. By inserting the string hypothesis \eqref{d23-strings} \emph{including the deviations} into the Bethe equations \eqref{bae00}, we multiply together the equations of complex conjugated pairs and then take the logarithm of the resulting equations, and introduce
	\begin{equation}
		x_{-j}=-x_j\,,\qquad\qquad y_{-j}=-y_j \,,
	\end{equation}
	to obtain: 
	\begin{equation}\label{fmdkmdkfmsöösmm}
		\begin{aligned}
			2 I^{x}_j=&+\frac{2N}{\ri \pi} \ln\left(\frac{\cosh(\ri \delta_j-x_j)}{\cosh(\ri \delta_j+x_j)}\right)+\frac{2N}{\ri \pi}\ln\left( \frac{\cosh(\ri \delta_j+2\ri \gamma+x_j)}{\cosh(\ri \delta_j+2\ri \gamma-x_j)} \right)\\
			&+\textcolor{red}{\frac{1}{\ri\pi}\ln\left(\frac{\sinh(2\ri \delta_j-2x_j)}{\sinh(2\ri \delta_j+2x_j)} \right)} +\frac{1}{\ri\pi}\ln\left(\frac{\sinh(2\ri \delta_j+4\ri \gamma+2x_j)}{\sinh(2\ri \delta_j+4\ri \gamma-2x_j)} \right)\\
			&-\frac{1}{\ri \pi}\textcolor{red}{\sum^{\frac{m_1}{2}}_{k=-\frac{m_1}{2}}\left\{\ln\left(\frac{\sinh(\ri (\delta_j+\delta_k)-(x_j-x_k))}{\sinh(\ri (\delta_j+\delta_k)+(x_j-x_k))}\right. \right)}+\ln\left(\frac{\sinh(\ri (\delta_j+\delta_k)+4\ri \gamma+(x_j-x_k))}{\sinh(\ri (\delta_j+\delta_k)+4\ri \gamma-(x_j-x_k))} \right)\\
			&+\left.\ln\left(\frac{\sinh(2\ri \gamma-\ri (\delta_j-\delta_k)+(x_j-x_k))}{\sinh(2\ri \gamma-\ri (\delta_j-\delta_k)-(x_j-x_k))} \right)+\ln\left(\frac{\sinh(2\ri \gamma+\ri (\delta_j-\delta_k)+(x_j-x_k))}{\sinh(2\ri \gamma+\ri (\delta_j-\delta_k)-(x_j-x_k))} \right)\right\}\\
			&-\frac{1}{\ri \pi}\sum^{\frac{m_2}{2}}_{k=-\frac{m_2}{2}}2\ln\left(\frac{\sinh(2\ri\gamma+\ri \delta_j-(x_j-x_k))}{\sinh(2\ri\gamma+\ri \delta_j+(x_j-x_k))} \right)+\textcolor{red}{2\ln\left(\frac{\sinh(\ri \delta_j+(x_j-y_k))}{\sinh(\ri \delta_j-(x_j-y_k))} \right)}\,,
		\end{aligned}
	\end{equation}
	and 
	\begin{equation}
		\begin{aligned}
			I^y_j=&\frac{1}{\ri\pi} \ln\left(\frac{\cosh(\ri \gamma+y_j)}{\cosh(\ri \gamma-y_j)}\right)+\frac{1}{\ri\pi}\sum^{\frac{m_1}{2}}_{k=-\frac{m_1}{2}}\,\ln\left(\frac{\sinh(2\ri \gamma+\ri \delta_k+y_j-x_k)}{\sinh(2\ri \gamma+\ri \delta_k-y_j+x_k)}\right)+\textcolor{red}{\ln\left(\frac{\sinh(\ri \delta_k+x_k-y_j)}{\sinh(\ri \delta_k-x_k+y_j)}\right)}\\
			&+\frac{1}{\ri\pi}\sum^{\frac{m_2}{2}}_{k=-\frac{m_2}{2}}\ln\left(\frac{\sinh(2\ri\gamma-y_j+y_k)}{\sinh(2\ri\gamma+y_j-y_k)}\right) \,,
		\end{aligned}
	\end{equation}
	where $I^{x,y}_k$ are half integers for $m_{1,2}$ even. Note that the red terms would not be present if one blindly inserts the string hypothesis without the deviation into the Bethe equations, as they would cancel out if one multiplies out the strings.   
	Taking now the limit $\delta_j\to 0$ and using that 
	\begin{align}
		\theta(x)=\lim_{\delta\to 0}\,\frac{1}{\ri \pi}\, \ln\left(\frac{\sinh(\ri \delta-x)}{\sinh(\ri \delta+x)}\right) \,,
	\end{align}
	we obtain
	\begin{equation}
		\begin{aligned}
			2 I^x_j=&+\frac{2N}{\ri \pi}\ln\left(\frac{\cosh(2\ri \gamma+x_j)}{\cosh(2\ri \gamma-x_j)} \right)+\theta(x_j) +\frac{1}{\ri\pi}\ln\left(\frac{\sinh(4\ri \gamma+2x_j)}{\sinh(4\ri \gamma-2x_j)} \right)\\
			&-\frac{1}{\ri \pi}\sum^{\frac{m_1}{2}}_{k=-\frac{m_1}{2}}\ri \pi\,\theta(x_j-x_k)+\ln\left(\frac{\sinh(4\ri \gamma+(x_j-x_k))}{\sinh(4\ri \gamma-(x_j-x_k))} \right)+2\ln\left(\frac{\sinh(2\ri \gamma+(x_j-x_k))}{\sinh(2\ri \gamma-(x_j-x_k))} \right)\\
			&-\frac{1}{\ri \pi}\sum^{\frac{m_2}{2}}_{k=-\frac{m_2}{2}}2\ln\left(\frac{\sinh(2\ri\gamma-(x_j-x_k))}{\sinh(2\ri\gamma+(x_j-x_k))} \right)-2\ri \pi\,\theta(x_j-y_k)\,,\\
			I^y_j=&\frac{1}{\ri\pi} \ln \left( \frac{\cosh(\ri\gamma + y_j)}{\cosh(\ri\gamma - y_j)} \right)
			+  \frac{1}{\ri\pi} \sum^{\frac{m_1}{2}}_{k=-\frac{m_1}{2}} \ri\pi\, \theta(y_j - x_k)
			+ \ln \left( \frac{\sinh(2\ri\gamma - x_k + y_j) }{\sinh(2\ri\gamma + x_k - y_j) } \right)\\
			&+\frac{1}{\ri\pi}\sum^{\frac{m_2}{2}}_{k=-\frac{m_2}{2}}\ln\left(\frac{ \sinh(2\ri\gamma + y_k- y_j)}{\sinh(2\ri\gamma - y_k + y_j)}\right) \,.
		\end{aligned}
	\end{equation}
	From this point onward, everything follows again the standard procedure: we introduce the monotonic increasing counting functions as $z_{\alpha}(\alpha_j)=\frac{I^{\alpha}_j}{N}$ for $\alpha=x,y$. The root density is obtained by differentiating the corresponding counting function $\rho^\alpha=\frac{\rm d}{{\rm d}\alpha} z^{\alpha}(\alpha)$. By using $\theta'(x)=\delta(x)$ and applying the Euler-Maclaurin formula to approximate sums by integrals, one obtains in the limit $N \to \infty$ a linear integral equation with infinite boundaries:
	\begin{align*}
		3\rho^{x}(x) - 2\rho^{y}(x) &= \frac{2\sin(4\gamma)}{\pi\left( \cos(4\gamma) + \cosh(2x) \right)} - \frac{1}{N}\frac{2\sin(8\gamma)}{\pi\left( \cosh(4x) - \cos(8\gamma) \right)} + \frac{1}{N}\delta(x) \\
		&\quad + \int_{-\infty}^{+\infty} \mathrm{d}\tilde{x} \, \rho^{x}(\tilde{x}) \left( \frac{2\sin(4\gamma)}{\pi\left( \cosh(2\tilde{x} - 2x) - \cos(4\gamma) \right)} + \frac{\sin(8\gamma)}{\pi\left( \cosh(2\tilde{x} - 2x) - \cos(8\gamma) \right)} \right) \\
		&\quad - 2 \int_{-\infty}^{+\infty} \mathrm{d}\tilde{y} \, \rho^{y}(\tilde{y}) \left( \frac{\sin(4\gamma)}{\pi\left( \cosh(2x - 2\tilde{y}) - \cos(4\gamma) \right)} \right) \,,
	\end{align*}
	and
	\begin{align*}
		\rho^{y}(y) - \rho^{x}(y) &= \frac{1}{N}\frac{\sin(2\gamma)}{\pi\left( \cos(2\gamma) + \cosh(2y) \right)} - \int_{-\infty}^{+\infty} \mathrm{d}\tilde{x} \, \rho^{x}(\tilde{x}) \frac{\sin(4\gamma)}{\pi\left( \cosh(2\tilde{x} - 2y) - \cos(4\gamma) \right)} \\
		&\quad + \int_{-\infty}^{+\infty} \mathrm{d}\tilde{y} \, \rho^{y}(\tilde{y}) \frac{\sin(4\gamma)}{\pi\left( \cosh(2\tilde{y} - 2y) - \cos(4\gamma) \right)} .
	\end{align*}
	Using the Fourier transform, these equations can be solved order-by-order in the system size $N$ by setting $\rho^\alpha=\sigma^\alpha+\frac{1}{N}\tau^\alpha$ , yielding $\sigma^x=\sigma^y$ and $\tau^x$ as given by \eqref{fmvkmcvncvnjs} and \eqref{fmdkfdkfdknfjn}.

\providecommand{\href}[2]{#2}\begingroup\raggedright\endgroup

\end{document}